\documentclass[12pt]{article}
\usepackage{amsmath}
\usepackage{graphicx}
\usepackage{enumerate}
\usepackage{natbib}
\usepackage{url} 
\usepackage{booktabs}
\usepackage[table]{xcolor}
\usepackage{multirow}
\usepackage{lscape}
\usepackage{times}

\newcommand{\blind}{1}

\usepackage{probstat}
\RequirePackage{amsthm,amsmath,amsfonts,mathtools}
\usepackage{xr-hyper}
\usepackage{hyperref}
\hypersetup{colorlinks,citecolor=blue,urlcolor=blue}
\makeatletter

\newtheorem{assumption}{Assumption}
\usepackage{apptools}
\AtAppendix{\counterwithin{assumption}{section}}
\allowdisplaybreaks
\usepackage{authblk}

\begin{document}

\def\spacingset#1{\renewcommand{\baselinestretch}%
{#1}\small\normalsize} \spacingset{1}

\date{}
\if1\blind
{
  \title{\bf A Spatial Interference Approach to Account for Mobility in Air Pollution Studies with Multivariate Continuous Treatments}
  \author[1]{Heejun Shin\thanks{
      Research described in this article was conducted under contract to the Health Effects Institute (HEI), an organization jointly funded by the United States Environmental Protection Agency (EPA) (Assistance Award No. CR-83590201) and certain motor vehicle and engine manufacturers. The contents of this article do not necessarily reflect the views of HEI, or its sponsors, nor do they necessarily reflect the views and policies of the EPA or motor vehicle and engine manufacturers. The authors would like to thank Georgia Papadogeorgou and Corwin Zigler for insightful comments on the proposed methodology. The authors would like to thank Cuebiq for both access to mobility data, as well as their willingness to help in all technical aspects regarding this data source. The computations in this paper were run on the FASRC Cannon/FASSE cluster supported by the FAS Division of Science Research Computing Group at Harvard University. Both key data sources used in this manuscript are not able to be made publicly available. The Medicare records used in this study are obtained through the Research Data Assistance Center (ResDAC; \url{https://resdac.org/}). Access requires a CMS-approved research request and an executed Data Use Agreement; for identifiable files, IRB approval is also required. ResDAC documents eligibility, required materials, and available files. Additionally, the cell phone mobility data was provided by Cuebiq through an agreement with the University of Florida, and only certified users have access to this proprietary data. }}
    \author[1,2]{Danielle Braun}
    \author[1]{Kezia Irene}
    \author[1]{Michelle Audirac}
    \author[3]{Joseph Antonelli}
    \affil[1]{Department of Biostatistics, Harvard T.H. Chan School of Public Health, Boston, MA}
    \affil[2]{Department of Data Science, Dana-Farber Cancer Institute, Boston, MA}
    \affil[3]{Department of Statistics, University of Florida, Gainesville, FL}
  \maketitle
} \fi

\if0\blind
{
  \bigskip
  \bigskip
  \bigskip
  \begin{center}
    {\LARGE\bf A Spatial Interference Approach to Account for Mobility in Air Pollution Studies with Multivariate Continuous Treatments}
\end{center}
  \medskip
} \fi

\begin{abstract}
We develop new methodology to improve our understanding of the causal effects of multivariate air pollution exposures on public health accounting for mobility. Typically, in environmental health studies, exposure to air pollution for an individual is assigned based on their residential address, though many people spend time in different regions with potentially different levels of air pollution. To account for this, we incorporate estimates of the mobility of individuals from cell phone mobility data to obtain a more accurate estimate of their air pollution exposure. We treat this as an interference problem, where individuals in one geographic region can be affected by exposures in other regions due to mobility into those areas. We propose policy-relevant estimands and derive expressions showing the extent of bias one would obtain by ignoring individuals' mobility. We additionally highlight the benefits of the proposed interference framework relative to a measurement error framework to account for mobility. Utilizing flexible Bayesian methodology we develop novel estimation strategies to estimate causal effects that account for this spatial spillover.  Lastly, we use the proposed methodology to study the health effects of ambient air pollution on mortality among Medicare enrollees in the United States.
\end{abstract}

\noindent%

\spacingset{1} 
\section{Introduction}

Estimating the health effects of air pollution is a crucially important scientific question due to the large public health burden of exposure to air pollution \citep{cohen2017estimates, burnett2018global}. In most environmental health studies, exposure to air pollution is measured at an individuals' residence or home geographic area. This ignores the fact that people spend time in different geographic regions with potentially different levels of air pollution, and not accounting for this can potentially lead to biased estimates of the effects of air pollution on health. Additionally, most work has focused on examining the impact of a single pollutant at a time. Humans are exposed to an array of ambient air pollutants, which has led to a shift towards estimating the health effects of multiple pollutants simultaneously \citep{dominici2010protecting}. In this work, we address both of these limitations by investigating the impact of mobility on the estimation of the health effects of air pollution in a way that can be incorporated with both single or multi-pollutant analyses. 

Estimating the causal effects of environmental pollutants comes with a number of challenges, some of which are ubiquitous when utilizing observational data, while others are unique to environmental or spatial settings \citep{reich2021review}. Unmeasured confounding is always a threat to observational studies, though some work has been done to alleviate this if the unmeasured variables are spatially correlated, which is common in environmental settings \citep{papadogeorgou2019adjusting}. When dealing with multiple pollutants, estimation of the health effects of air pollution also becomes difficult, which has led to the development of flexible approaches that allow for nonlinear associations and interactions between pollutants \citep{herring2010nonparametric,bobb2015bayesian, antonelli2020estimating}. Another challenge when estimating the causal effects of multivariate exposures is that it can be difficult to define policy-relevant estimands when pollutants are highly correlated and potentially emitted from the same source. Despite these challenges, a causal inference approach to environmental epidemiology has been advocated for \citep{carone2020pursuit, sommer2021assessing}. Reasons for this push include the clear definition of target estimands, removing reliance on a particular statistical model, the ability to use flexible approaches, clarity on the assumptions needed to identify causal effects, and sensitivity analyses to evaluate potential violations of these assumptions \citep{dominici2017best}. 

Another issue of particular relevance to this work is interference, which occurs when units are affected by the exposure status of other units in the study \citep{hong2006evaluating, sobel2006randomized, hudgens2008toward}. A critical issue when dealing with interference is simplifying the manner in which the exposure status of certain units can affect the outcomes of other units. Partial interference is commonly assumed, where interference is restricted to clusters or a subset of the data points \citep{vanderweele2011bounding, tchetgen2012causal, papadogeorgou2019causal}. Other approaches incorporate spatial proximity, where the treatment status of nearby observations are more likely to affect a unit's outcome \citep{wang2020design, giffin2022generalized}. More general interference structures, such as those seen in social networks can also be accommodated \citep{aronow2017estimating, ogburn2022causal}. These approaches have also been used in the air pollution literature to better understand the impact of interventions on power plants \citep{zigler2020bipartite, zigler2021bipartite}, or the impacts of the EPA's non-attainment designations \citep{zirkle2021addressing}. Recently, there have been efforts to investigate unknown or misspecified interference structures \citep{savje2021average, weinstein2023causal}. Despite this surge in interest, nearly all work has focused on univariate, binary treatments.

In this paper we address a number of important issues to understand the impact of air pollution on health outcomes holistically. Our approach estimates causal effects for both single and multiple pollutant studies, while simultaneously accounting for mobility of individuals across geographic regions. We address mobility by incorporating cell phone mobility data and a novel interference framework. Specifically, we let outcomes depend on exposures at both 1) the residential geographic region of individuals, and 2) a weighted average of exposure at neighboring areas based on weights determined by cell phone mobility data. This extends the literature on causal inference under interference to settings with multivariate, continuous treatments, while leading to improved estimates of the overall impact of air pollution on health. We develop policy-relevant estimands for causal effects of the air pollution mixture, and derive bias formulas for when mobility is ignored, as is typically the case in environmental health epidemiology. Further, we develop flexible Bayesian causal estimators in this setting, and use our framework to gain epidemiological insights on the effects of air pollution on mortality in the elderly United States population. 

\section{Potential outcomes and estimands}\label{sec: setting}

In our motivating study, the unit of analysis is the zip code, and the mobility data, originally measured at the individual level, are aggregated to form zip-code-level summaries that define the network of population movement. Specifically, we consider an $n\times n$ adjacency matrix $\bT$ where the $(i,j)-$th entry, $T_{ij}$, represents the total amount of time cell phone users residing in zip code $i$ spent in zip code $j$. The $i$-th row of this matrix, denoted by $\bT_i=(T_{i1},...,T_{in})$, therefore, summarizes the outgoing mobility information of zip code $i$ to all zip codes. Aggregated mobility data is provided by Cuebiq, a location intelligence platform. Data is collected from anonymized users who have opted-in to provide access to their location data anonymously, through a CCPA and GDPR-compliant framework. Through its Social Impact program, Cuebiq provides mobility insights for academic research and humanitarian initiatives. The Cuebiq responsible data sharing framework enables research partners to query anonymized and privacy-enhanced data, by providing access to an auditable, on-premise Data Cleanroom environment. All final outputs provided to partners are aggregated in order to preserve privacy.

Further, we observe $\boldsymbol{D}_i = (Y_i, \boldsymbol{W}_i, \boldsymbol{X}_i)$ for $i=1, \dots, n$. $Y_i$ is an outcome of interest for zip code $i$. We let $\boldsymbol{W}_i$ denote the $q$-dimensional vector of exposures, while $\boldsymbol{X}_i$ is a $p$-dimensional vector of pre-treatment confounders. Throughout we let lowercase letters represent realizations of random variables. In the following, we frequently omit the subscript $i$ for ease of exposition.

\subsection{Potential outcomes under interference}\label{sec: assumptions}

It is commonly assumed that potential outcomes of a unit are not affected by exposure levels of other units. Typically this assumption is referred to as \textit{no interference} between units \citep{cox1958planning} or, when combined with the \textit{no-multiple-versions-of-treatment} assumption, the stable unit treatment value assumption (SUTVA, \citealp{rubin1980randomization}). To see how this assumption doesn't hold in our setting, consider two scenarios in which a unit is exposed to the same level of exposure in its own geographic area, but is exposed to either low or high levels of exposure during day-to-day mobility. If these exposures have an effect on the outcome, we would expect the potential outcomes in these two scenarios to differ, leading to violations of SUTVA due to multiple versions of treatment that depend on the level of neighboring exposure values.  For this reason, we first define potential outcomes as $Y_i(\boldsymbol{w}^{all}; \bT_i)$, which is the outcome we would observe for unit $i$ if the entire sample were exposed to global pollution levels $\boldsymbol{w}^{all} = (\bw_1', \bw_2',...,\bw_n')'$ with the mobility network $\bT_i$, and assume that $Y_i = Y_i(\bW^{all}; \bT_i)$. We note that $\bT_i$ is not a primary exposure of interest (e.g. air pollution) but rather a given structural feature governing interference. To distinguish this from primary exposures, we use a semicolon in the potential outcomes notation. This general formulation leads to a prohibitively large set of potential outcomes, since each unit’s outcome could, in principle, depend on the exposures and mobility patterns of all other units. 

Before proceeding it is important to note that while we choose to formulate this as a problem of interference, one could alternatively view it as one of measurement error, where the exposure within a zip code is a mismeasured version of the true exposure to pollution. Regardless of whether it is viewed as a problem of interference or measurement error, the key issue is that the standard SUTVA assumption is violated and potential outcomes depending on only a zip code's own exposure are not well-defined. Mobility information could also be used to help correct for measurement error, but we choose to utilize an interference framework for multiple reasons. For one, many of the techniques we adopt, such as exposure mappings in the following section, are derived from, and easiest to understand, from an interference framework. Additionally, we show in Section \ref{sec:Bias} that the interference framework is slightly more general when the effects of air pollution are different when traveling to other zip codes, which is possible if individuals spend more time inside or outside while traveling compared with when they're in their home zip code. 

\subsection{Exposure mapping using cell phone mobility data}\label{sec:mobility}

\begin{figure}[t]
    \centering
    \includegraphics[width=0.95\linewidth]{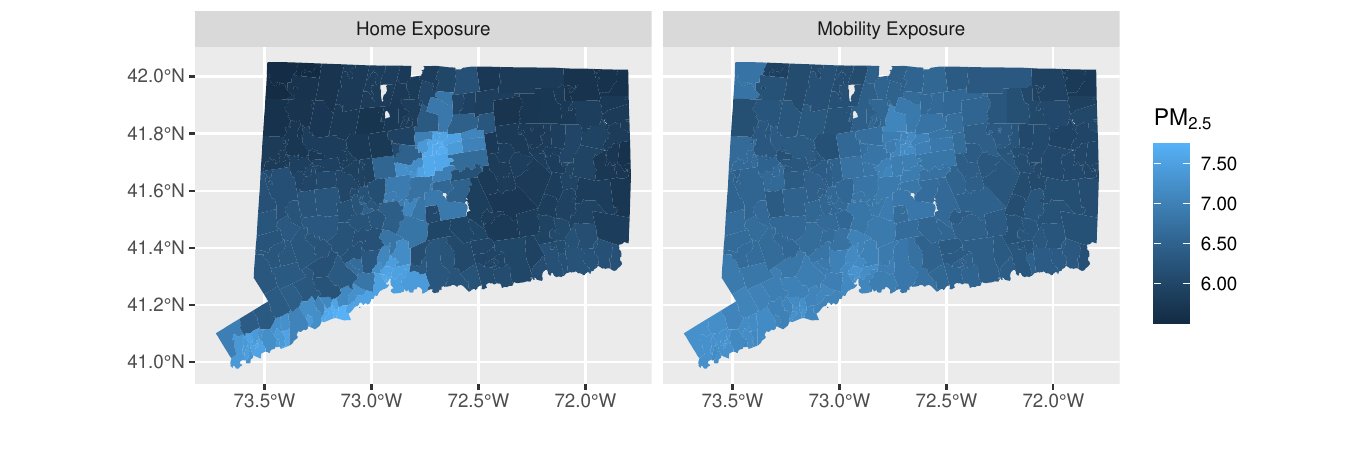}
    \caption{Residential and mobility-based exposure to \pmtpfs in Connecticut in 2016.}
    \label{fig:BCmobility}
\end{figure}

To reduce the complexity of potential outcomes, we adopt the exposure mapping framework \citep{aronow2017estimating, forastiere2021identification}, where potential outcomes are assumed to depend only on $d_i(\bw^{all}; \bT_i)$, a lower-dimensional summary of the global exposure matrix $\bw^{all}$ and the unit's outgoing network $\bT_i$. Specifically, we define an exposure mapping as $d_i(\bw^{all}; \bT_i) = (\bw_i, \bg_i; \tau_i)$ where $\bw_i$ is unit $i$’s own exposures, $\bg_i = g(\bw^{all}_{-i}; \bT_i)$ represents a combined summary of other units' exposures using unit $i$'s outgoing network, and $\tau_i = k(\bT_i)$ is a summary of unit $i$'s outgoing network. For each $i$, we define $\tau_i = T_{ii} / \sum_{k} T_{ik},$ which is the proportion of time that cell phones from zip code $i$ spent in their residential zip code. Next, define $\boldsymbol{\alpha}_i$ to be an $n-$dimensional vector of weights such that $\alpha_{ii} = 0$ and $\alpha_{ij} = T_{ij}/\sum_{k \neq i} T_{ik}$ for $i \neq j$. These weights reflect the proportion of the time spent outside of their residential zip code that was spent in zip code $j$. We define mobility-based exposure as $\bg_i = \sum_{j=1}^n \alpha_{ij} \bw_j,$ which is a weighted average of exposure in other zip codes weighted by the amount of time spent in each zip code. Figure \ref{fig:BCmobility} shows both residential and mobility-based exposures for annual average \pmtpfs exposure in Connecticut in 2016. The left panel shows that exposure is elevated along the main interstate in Connecticut as well as near the largest city, Hartford. The right panel shows that other regions are also exposed to these elevated \pmtpfs levels due to their daily travel patterns. 

We formalize the connection between the exposure mapping $d_i(\bw^{all}; \bT_i) = (\bw_i, \bg_i; \tau_i)$ and potential outcomes under interference as follows:
\begin{assumption}\label{asmp: Exposure mapping}
    For any $\bw^{all}$ and ${\bw^{all}}'$, we have $Y_i(\bw^{all};\bT_i) = Y_i({\bw^{all}}';\bT_i)$ whenever $(\bw_i, \bg_i; \tau_i) = (\bw'_i, \bg'_i; \tau_i)$.
\end{assumption}
\noindent This assumption implies that, once the network is given, the potential outcome for zip code $i$ depends on its own exposure $\bw_i$, and all remaining exposures, but only through $g(\cdot)$. Different distributions of neighboring exposures are assumed to have the same spillover effect as long as their mobility-weighted averages are the same. While other weighting schemes could be used based on prior knowledge of exposure relevance or biological mechanism, we believe that our choice of $g(\cdot)$, which weights the neighboring exposures by the proportion of time spent in each location, represents a reasonable and interpretable default. Note also that we make no assumptions about the mobility patterns of individuals within a zip code, but instead assume that the only relevant quantity for our zip code level analysis is their aggregate mobility patterns. This simplification substantially reduces the potential outcome space and allows us to write our potential outcomes as $Y_i(\boldsymbol{w}, \boldsymbol{g};\tau_i)$, which is the outcome we would observe if zip code $i$ was exposed to residential exposure levels $\boldsymbol{w}$ for a proportion $\tau_i$ of time and mobility-based exposure levels $\boldsymbol{g}$ for the remaining $1-\tau_i$. For notational simplicity, we omit $\tau_i$ from the potential outcomes when it is clear from context, with the understanding that they inherently depend on $\tau_i$.

\subsection{Policy-relevant estimands}\label{sec: estimands}

Now that we have defined potential outcomes, we can discuss the estimands of interest. Certain zip codes have different exposure values with differential impacts of a policy aimed at reducing air pollution. For instance, implementing a policy at power plants would not reduce exposure in areas of the country far from a power plant. To accommodate this, we define a sample-level estimand with heterogeneous changes in exposure across zip codes. We let the shift in exposures across all $n$ zip codes be represented by $\bDelta = [\bDelta_1', \dots, \bDelta_n']'$, where $\bDelta_i \in \reals^q$ denotes the change in zip code $i$. These shifts lead to corresponding changes in both residential exposure, $\bDelta_{wi} = \bDelta_i$, and mobility-based exposure $\bDelta_{gi}$, which is computed using a known exposure mapping function that accounts for mobility patterns. Specifically, for zip code $i$, the new mobility-based exposure is $g(\bw^{all}_{-i} + \bDelta_{-i}; \bT_i) = \bG_i + \bDelta_{gi}$. We define the sample-level estimand as:
\begin{align*}
    \omega(\bDelta) &= \frac{1}{n} \sum_{i=1}^n \bigg\{ Y_i(\bw_i + \bDelta_{wi}, \bg_i + \bDelta_{gi}) - Y_i(\bw_i, \bg_i) \bigg\}
\end{align*}
This estimand represents the average effect of applying a policy-induced exposure shift, which may vary across regions. This estimand is closely related to the stochastic interventions proposed in \cite{haneuse2013estimation} that shift exposures from their natural levels. Note that the focus on shifted interventions makes the positivity assumption more plausible than for estimands focusing on constant or extreme exposure levels. This also has connections to the incremental effect framework \citep{bonvini2023incremental, schindl2024incremental}, which defines causal effects under stochastic interventions by tilting the generalized propensity score. These do not rely on positivity assumptions at all and therefore may be useful when positivity is a serious concern, although they have not been extended to accommodate multivariate exposures or interference. Additionally, we must distinguish our assignment-conditional estimands from marginal estimands commonly used in the interference literature \citep{hudgens2008toward, savje2021average}. These estimands average over the distribution of neighboring exposures and are often tied to hypothetical randomization schemes. In contrast, our estimand is defined conditional on observed or policy-induced exposure assignments, making them especially relevant for environmental health policy, where exposure changes are heterogeneous and spatially targeted, and the goal is to quantify health impacts under specific regulatory scenarios.

At times it will also be of interest to define an analogous sample-level group average treatment effect (GATE) for any subgroup $S\subset\{1,...,n\}$ with cardinality $n_s$ as
\begin{align*}
    \omega_s(\bDelta) = \frac{1}{n_s} \sum_{i\in S} \big\{ Y_i(\bw_i + \bDelta_{wi}, \bg_i + \bDelta_{gi}) - Y_i(\bw_i, \bg_i) \big\}.
\end{align*}
This estimand represents the average effect of the exposure shift $\bDelta$ that changes exposures for all zip codes, while the average is taken only over the subgroup $S$. Examining this estimand allows us to assess whether certain subgroups are more impacted by the exposure shift, and will provide information about which covariate profiles are most impacted by shifts in air pollution exposures. Lastly, as we have observational data, we must make additional assumptions to estimate these quantities from observed data. In particular, we make a conditional exchangeability assumption:
\begin{assumption}[Joint unconfoundedness]\label{asmp: unconfound}
$$Y(\boldsymbol{w}, \boldsymbol{g}) \independent \boldsymbol{W}, \boldsymbol{G} \mid \boldsymbol{X}$$
\end{assumption} 
\noindent This assumption states that we have measured all common causes of the exposures and the outcome $Y$. In the context of our air pollution application, this implies that there are no unmeasured zip-code-level characteristics that jointly affect both air quality and health outcomes. In other words, after adjusting for aggregated covariates such as demographic composition, socioeconomic status, and other environmental factors, differences in air pollution exposures (both residential and mobility-based) across zip codes can be considered as-if random with respect to potential outcomes. Although this assumption is not directly testable, it is made more plausible by including a rich set of covariates. One can weaken this assumption if repeated observations for each unit are available, such as repeated observations over time. We explore this possibility in Section \ref{sec:Differencing} where we describe an approach that is robust to time-invariant unmeasured confounding.
We also make a positivity assumption, which for our sample-level estimand is 
\begin{align*}
    f_{\bW,\bG|\bX}((\boldsymbol{W}_i + \boldsymbol{\Delta}_{wi}, \boldsymbol{G}_i+ \boldsymbol{\Delta}_{gi}) | \bX = \bX_i) > 0 \hspace{0.2 in}\text{for}\hspace{0.2 in}i=1,...,n.
\end{align*}
where $f_{\bW,\bG|\bX}(\cdot)$ is the conditional density of the exposures.
This is a weaker assumption than positivity assumptions for estimands that fix treatment at the same value for all units, particularly for small values of the intervention, because we are guaranteed that $f_{\bW,\bG|\bX}((\boldsymbol{W}_i, \boldsymbol{G}_i) | \bX = \bX_i) > 0$ by the presence of data point $i$.

\section{Bias when mobility is ignored}
\label{sec:Bias}

Here we derive expressions for the bias in estimating the impact of a policy to reduce air pollution that one would obtain if they ignored mobility into nearby regions. 

\subsection{Linear model}\label{sec: linear}
To provide intuition, we focus on a simplified setting with univariate exposures scaled to have marginal variance 1 and assume a linear model for the potential outcomes:
$$Y(w,g) = \tau w \beta_w + (1-\tau) g \beta_g + \epsilon,$$
where $\tau$ is the proportion of time spent in the residential zip code. Suppose the goal is to estimate the average effect of increasing pollution everywhere by one unit, which corresponds to $\tau \beta_w + (1 - \tau) \beta_g$. Two potential approaches are: (i) regressing the outcome on home exposure W only, and (ii) regressing on the time-weighted exposure $W^* = \tau W + (1 - \tau) G$. The first approach ignores mobility, which is the standard approach when studying health effects of pollution, and estimates the effect to be $\tau \beta_w + \rho (1 - \tau) \beta_g$, where $\rho$ is the correlation between $W$ and $G$. Note this is biased unless $\rho=1$, although importantly when $\beta_w$ and $\beta_g$ have the same sign, which is reasonable in environmental applications, this approach will underestimate the effect of pollution. The second approach presents an alternative way of incorporating mobility information that falls more within the measurement error framework by providing a correctly measured exposure $W^*$. We derive the exact form of the causal effect obtained using the second approach in Appendix \ref{sec:BiasProof}, though it is generally closer to the truth than the first approach that ignores mobility. However, we show that it still yields biased estimates when $\beta_w \neq \beta_g$, i.e., when health effects of the residential and mobility-based exposures differ. One reason this could occur is if people spend more time inside or outside while in their home zip code, as pollution exposure can differ substantially between indoor and outdoor exposure. This shows the importance of separately modeling the effects of $W$ and $G$ as it provides unbiased estimates of health effects regardless of the value of $(\tau, \rho, \beta_w, \beta_g)$. Full derivations and discussion are provided in Appendices \ref{sec:appendix_BiasLinear} and \ref{sec:BiasProof}. 

\subsection{General bias formula}\label{sec: general bias}
Now we investigate the bias incurred by ignoring mobility when estimating $\omega(\bDelta)$ without assuming univariate exposures or a linear model. Let $m(\bw,\bg,\bx;\tau)=\E(Y \vert \bW=\bw, \bG=\bg, \bX=\bx; \tau)$ denote the true outcome regression function, and define the oracle plug-in estimator that correctly adjusts for both residential and mobility-based exposures as
\begin{align*}
    \omega_{oracle}(\bDelta) = \frac{1}{n} \sum_{i=1}^n \big\{ m(\bw_i + \bDelta_{wi}, \bg_i + \bDelta_{gi},\bx_i;\tau_i) - Y_i \big\}
\end{align*}
which satisfies $\E[\omega_{oracle}(\bDelta) - \omega(\bDelta)]=0$. Suppose instead that the outcome model is misspecified by ignoring $\bG$ and $\tau$, denoted as $\widetilde{m}(\bw,\bx)$, and define the corresponding estimator
\begin{align*}
    \widetilde{\omega}(\bDelta) &=\frac{1}{n} \sum_{i=1}^n \big\{ \widetilde{m}(\bw_i + \bDelta_{wi}, \bx_i) - Y_i \big\}.
\end{align*}
It is straightforward to observe that the resulting difference $\widetilde{\omega}(\bDelta)-\omega_{oracle}(\bDelta)$, is given by
{\small
    \begin{align*}
        \dfrac{1}{n}\sumio\Big\{\int m(\bw_i+ \bDelta_{wi},\bg',\bx_i;\tau')dF_{G,\tau\vert W=\bw+ \bDelta_{wi},X=\bx_i}(\bg',\tau') - m(\bw_i + \bDelta_{wi}, \bg_i + \bDelta_{gi},\bx_i;\tau_i)\Big\}.
    \end{align*}
}
This result shows that we are only guaranteed to obtain unbiased estimates of $\omega(\bDelta)$ when  the mobility-related components ($\bG$ and $\tau)$ have no effect on $Y$, or when they are deterministic functions of $\bX$ and $\bW$. The reason for this bias is that the model which does not account for mobility, effectively averages over the range of possible values for mobility-related variables, instead of fixing them at the desired level of $\boldsymbol{g}_i + \bDelta_{gi}$ and $\tau_i$.

\section{Misspecification of weights}
\label{sec:MissWeights}

In this section we assess the robustness of our findings to misspecification of $\tau_i$, which is important because 1) the population from the cell phone mobility data may not be representative of the population that we are estimating the causal effect for, and 2) the values of $\tau_i$ are estimates from cell phone mobility data and there is inherent variability in these estimates. Regarding the first of these two concerns, although demographic information is not collected on the anonymous cell phone users, recent studies using the same source of mobility data have attempted to infer demographic characteristics of the population of cell phone users by linking cell phone data to publicly available census tract data on such characteristics. They have found that the users within the cell phone mobility data are largely representative of the United States population as a whole, as indicated by high correlations between their data and census or county level data \citep{deng2021network}. Despite this, it is important to study robustness of our findings to any discrepancies between the population of interest and the mobility population.

For simplicity of calculations, we use the scenario of Section \ref{sec: linear} and let both $W$ and $G$ be univariate random variables with correlation $\rho$. Throughout we denote the true values as $\tau_i^*$ and denote its estimate from data as $\tau_i$. The true outcome in this setting is generated by the following
\begin{align}
    Y_i = \beta_w^* \tau_i^* W_i + \beta_g^* (1 - \tau_i^*) G_i + \epsilon_i, \label{eqn:TrueModelMiss}
\end{align}
and our goal is to estimate $\omega(\bDelta)$ with $\bDelta = \Vec{1}_n$. In this setting, the target estimand is $\beta_w^*\widebar{\tau^*} + \beta_g^* (1 - \widebar{\tau^*})$ where $\widebar{\tau^*} = \sumio \tau_i^*/n$. For brevity, we denote the estimand by $\omega$ and its estimator using misspecified weights by $\widehat{\omega}=\widehat{\beta}_w\widebar{\tau} + \widehat{\beta}_g (1 - \widebar{\tau})$, where $\widehat{\beta}_w$ and $\widehat{\beta}_g$ are obtained from fitting model \ref{eqn:TrueModelMiss} with $\tau_i$ instead of $\tau_i^*$. 

\subsection{Misspecification by scalar multiplication}\label{sec: miss_by_scalar}
We first study misspecification of the form $\tau_i = \tau_i^*/c$ for all $i$ and $c >0$. This scenario can occur when the target population differs from the one used to measure mobility. We focus on this situation specifically, because in Section \ref{sec: data analysis} we estimate causal effects in the Medicare population in the United States, which includes individuals over 65 years of age, who are likely to travel less than the overall population observed in the cell phone mobility data. Assuming further that $\tau$ is independent of exposure levels, we show in Appendix \ref{sec: miss_bias_proof} that the asymptotic bias is given by
\begin{align}
    \lim_{n \rightarrow \infty} \E(\widehat{\omega}-\omega)=\frac{\beta_g^* (1 - c) (1 - \rho) \text{Var}(\tau) \Big[ \rho \E(\tau) - (1 + \rho) \E(\tau^2) \Big]}{\Big(\E(\tau^2) \E[(1 - \tau)^2] - \rho^2 \E^2[\tau (1 - \tau)] \Big)}, \label{eqn:TauCerror}
\end{align}
where the expectation is with respect to the distribution of $Y$ given $W$ and $G$. The bias depends on 1) the degree of misspecification, given by $1-c$, 2) the correlation between $W$ and $G$, and 3) the distribution of $\tau$. The bias will be close to zero if $c$ is near 1 and we have very little misspecification, or when $\rho$ is large. Lastly, the magnitude of bias depends on the variability in $\tau$ (and therefore $\tau^*$). In the extreme case where individuals spend equal amounts of time in their residential zip code and $\var(\tau^*)=0$, our estimate is unbiased. When travel weights are misspecified by 50\% $(0.5<c<1.5)$ and $\rho$ is greater than 0.4, which is realistic for environmental applications, the resulting absolute bias is below $0.05\beta_g^*$. This small bias under substantial misspecification suggests that effect estimates are moderately robust to this form of misspecification. Because $\rho$ and the moments of $\tau$ can be estimated from the observed data, the degree of bias can also be quantified empirically. For the Medicare data, we found that for $c = 0.5$ and $c=1.5$, the absolute bias was below $0.004\beta_g^*$, suggesting a high degree of robustness for our application of interest.

\subsection{Misspecification due to measurement error}\label{sec: misspecification measurement error}

Now we consider a scenario where $\tau^*_i=\tau_i+\eta_i\in[0,1]$ and $\eta_i$ are independent and identically distributed random variables having a mean of 0 and a finite second moment. In Appendix \ref{sec: miss_bias_proof}, we show that this leads to asymptotically unbiased estimation. A more realistic scenario is one where $\tau_i=\tau_i^*+\eta_i$, since the mobility estimates are taken from a finite sample size of cell phone users, and therefore exhibit variability around their true values. For simplicity, assume $W$ and $G$ are independent, which produces the largest bias in Section \ref{sec: linear} unless  $W$ and $G$ are negatively correlated, which is unlikely in practice. Given that this is essentially a worst case scenario, the bias expression presented below can be used as a bound of the bias we would get when the mobility weights are measured with noise. As shown in Appendix \ref{sec: miss_bias_proof}, $\E(\widehat{\omega} - \omega)$ converges to
{\small
\begin{align*}
    &\beta_w^*\left\{\frac{\E[(1-\tau^*-\eta)^2]\E[{\tau^*}^2]}{\E[(\tau^*+\eta)^2]\E[(1-\tau^*-\eta)^2]}\E(\tau^*)-\E(\tau^*)\right\} + \\
    &\qquad\beta_g^*\left\{\frac{\E[(\tau^*+\eta)^2]\E[(1-\tau^*)^2]}{\E[(\tau^*+\eta)^2]\E[(1-\tau^*-\eta)^2]}\E(1-\tau^*)-\E(1-\tau^*)\right\}\\
    &\equiv\beta_w^*\left\{\xi_w\E(\tau^*)-\E(\tau^*)\right\} + \beta_g^*\left\{\xi_g\E(1-\tau^*)-\E(1-\tau^*)\right\}.
\end{align*}
}
We show in Appendix \ref{sec: miss_bias_proof} that both $\xi_w$ and $\xi_g$ are continuous, decreasing functions of $\E(\eta^2)\geq 0$ with $\lim_{\E(\eta^2)\rightarrow\infty}\xi_w=\lim_{\E(\eta^2)\rightarrow\infty}\xi_g=0$ and positive second derivatives. If there is no measurement error, i.e., $\E(\eta^2) = 0$, then $\xi_w=\xi_g=1$, and we have unbiased estimates. To highlight a situation with significant amounts of measurement error, assume that $\tau^*$ follows a uniform distribution between 0.25 and 0.75, and the measurement error $\eta$ is uniformly distributed between -0.25 and 0.25. In this scenario, $\xi_w=\xi_g=0.93$, which leads to very small amounts of bias. A crucial finding from this result is that measurement error leads to conservative estimates of causal effects as long as the signs of $\beta_w^*$ and $\beta_g^*$ are the same, which is expected in practice. We also run a sensitivity analysis for our Medicare data analysis in Appendix \ref{app:ssec:MisspecifyMobility} by adding noise to both $\tau$ and $\boldsymbol{\alpha}$ and find that our results are relatively robust to such errors in mobility estimates.

\section{Estimation and inference}\label{sec: Estimation}

Our estimand described in Section \ref{sec: estimands} can be estimated using the conditional mean of the outcome, $\E(Y \mid \bW, \bG, \bX; \tau)$. We work within the Bayesian paradigm and obtain posterior samples of this quantity, denoted by $\E^{(b)}(Y \mid \bW, \bG, \bX; \tau)$ for $b=1,\dots,B$. Once these posterior samples are obtained, posterior samples of the sample-level estimand are computed as
\begin{align*}
    \omega^{(b)}(\bDelta) 
    &= \frac{1}{n} \sum_{i=1}^n \bigg\{
    \E^{(b)}(Y \mid \bW = \boldsymbol{W}_i + \bDelta_{wi},
    \bG = \boldsymbol{G}_i + \bDelta_{gi}, 
    \bX = \bX_i; \tau_i) - Y_i
    \bigg\},
\end{align*}
with posterior means and quantiles used for point estimation and inference.

We first consider a flexible outcome-regression estimator based on Bayesian additive regression trees (BART; \citealp{chipman2010bart}), which have been shown to be effective across a range of causal inference problems \citep{linero2022and}. BART is useful in this setting because the exposure-response surface may include nonlinearities, interactions among pollutants within $\boldsymbol{W}$ or $\boldsymbol{G}$, interactions between home and mobility-based exposures, or heterogeneity by covariates $\boldsymbol{X}$. To allow for these possibilities, we model $\E (Y | \bW, \bG, \bX; \tau)$ with a standard BART prior distribution (see \citealp{linero2017review} for a review), using $(\bW,\bG,\bX,\tau)$ as predictors.

We also consider a structured semiparametric estimator based on a basis expansion. In contrast to BART, this model explicitly encodes the mobility decomposition by allowing home and mobility-based exposure-response functions to be similar, but not necessarily identical. Specifically, we model
\begin{align*}
    \E (Y_i \mid \bW_i, \bG_i, \bX_i; \tau_i) 
    &= \bX_i^\top \boldsymbol{\theta} 
    + \sum_{j=1}^q \Big\{ 
    \tau_i \boldsymbol{\phi}_{w,ij}^\top \boldsymbol{\beta}_j 
    + (1 - \tau_i) \boldsymbol{\phi}_{g,ij}^\top(\boldsymbol{\beta}_j + \boldsymbol{\zeta}_j) 
    \Big\},
\end{align*}
where $\boldsymbol{\phi}_{w,ij}$ and $\boldsymbol{\phi}_{g,ij}$ are basis expansions of the home and mobility-based exposures for pollutant $j$. The shared coefficient vector $\boldsymbol{\beta}_j$ contributes to both exposure-response functions, while $\boldsymbol{\zeta}_j$ captures deviations of the mobility-based response from the home response. We place shrinkage priors on $\boldsymbol{\zeta}_j$ so that information can be borrowed across the two functions when their effects are similar, while still allowing them to differ. Full prior specifications and posterior computation details are provided in Appendix \ref{sec: full conditional derivation}. Simulations in Appendix \ref{sec:Simulation} show that this shrinkage improves estimation when home and mobility-based exposure effects are similar and does not negatively affect inference when these effects differ. They also show that, even under misspecified mobility parameters, accounting for mobility performs better than ignoring mobility completely.

Additionally, we propose a distinct approach to estimation that is based on inverse propensity score weighting (IPW). Recent work has shown that estimands that shift exposure from their natural, observed levels to a shifted value, such as $\omega(\boldsymbol{\Delta})$, can be reformulated as a standard binary treatment problem within an augmented data set \citep{jiang2025exploring}. We provide specific details of this strategy in Appendix \ref{app:BinaryTreatment}, though importantly this provides a completely distinct approach to estimating the causal effect that does not rely on outcome modeling or any distributional assumptions for the outcome, and leads to similar results in the application of interest. 

\subsection{Incorporating spatial correlation}
\label{ssec:SpatialCorr}

The previous sections have focused on estimation of the conditional mean of the outcome. Our uncertainty quantification requires a full data generating model for the outcome. Given exposures and covariates, outcomes are assumed normally distributed and independent across zip codes. It is possible, however, that nearby zip codes may exhibit spatial correlation. To address this, we explore incorporating spatial random effects through conditionally autoregressive prior distributions \citep{hodges2003precision}, which allow for neighboring zip codes to have conditionally dependent outcomes, and can be incorporated simultaneously with any of the proposed approaches for estimating the conditional mean of the outcome. Technical details of this prior specification are provided in Appendix \ref{app:SpatialICAR} along with Medicare analysis results incorporating spatial random effects. For simplicity, we assume the outcomes are conditionally independent in the main analyses, but as we see in Appendix \ref{app:SpatialICAR}, incorporating spatial correlation makes very little impact on the resulting point estimates or credible intervals.

We acknowledge a related line of work that has explored when valid uncertainty quantification is possible under spatial interference. They show that valid inference requires certain limitations on the degree of interference across units \citep{savje2021average, leung2022causal, wang2025design}. These insights, however, have all focused on design-based inference under randomized trials where uncertainty stems from the treatment assignment mechanism and therefore they do not apply to our observational study setting here where inference is model-based rather than design-based. Nonetheless, analysis of cell phone mobility data shows that most individuals travel to a limited number of other zip codes: on average, 95\% of travel is directed to just 2\% of all possible destinations. Mobility is therefore quite localized, supporting the assumption that interference is restricted to a moderately small set of neighbors.

\section{Causal effects of air pollution mixtures}\label{sec: data analysis}

Here we examine the causal effect of total \pmtpfs as well as five components of \pmtpfs (black carbon (BC), ammonium (NH$_4^+$), nitrates (NO$_3^-$), organic matter (OM), and sulfates (SO$_4^{2-}$)) on mortality rates for Medicare enrollees in the contiguous United States with 30,448 zip codes from 2000 to 2016. While our approach can be applied to single exposure analyses (see Appendix \ref{app:pm_analysis} for \pmtpfs only results), we mostly focus on multivariate exposures with the five exposures above as this offers a more comprehensive assessment of air pollution exposure. We obtain air pollution data on a ($0.01^\circ \times 0.01^\circ$) monthly grid from the Atmospheric Composition Analysis Group \citep{van2019regional} and aggregate them to the yearly level. Note that these exposure values are predictions taken from statistical models that incorporate information from ground-based air pollution monitors, chemical transport models, satellite measurements of aerosol optical depth, and other land-use covariates. While exposure predictions for this data and related data sources have been shown to have high correlation with true exposure values \citep{van2019regional,xiao2021evaluation}, there is inherent error involved with exposure assessment. Such exposure measurement error is well-studied in air pollution epidemiology \citep{sheppard2012confounding} and is shown to decompose into both a berkson-like and classical-like form of measurement error. While the bias from this error can be in either direction \citep{szpiro2013measurement}, it is typically towards the null value of no exposure effect, and recent measurement error corrections on similar analyses within the Medicare cohort have led to more pronounced effect estimates than ignoring measurement error \citep{wu2019causal}. For this reason we ignore this source of measurement error in our analysis, though future work could look to incorporate measurement error corrections simultaneously with the incorporation of mobility data as we perform here. 

We observe age, sex, race, and dual eligibility to Medicaid, a proxy for low socioeconomic status, for each beneficiary in a zip code, which are aggregated to the zip code level. We also have obtained average BMI, smoking rates, median household income, median house value, the proportion of residents in poverty, the proportion of residents without a high school diploma, population density, and percent owner-occupied housing from the United States Census Bureau and the Center for Disease Control's Behavioral Risk Factor Surveillance System. Summer and winter temperature and humidity records are obtained from the National Climatic Data Center. Our outcome of interest is the annual mortality rate in each zip code. Mobility data comes from over 15 million cell phone users in the contiguous United States over all days in 2019. Raw Medicare data were transformed into analysis-ready datasets using a reproducible, modular, five-stage pipeline: parsing, harmonization, normalization, cleaning, and materialization. Parsing converts the annual fixed-width CMS files into structured tables suitable for efficient out-of-core computation. Harmonization uses prespecified, year-specific specifications to standardize variable names, data types, and categorical codings across years, resolving year-to-year drift in the CMS releases and producing longitudinal data with a consistent schema. Normalization organizes the harmonized data into relational tables for beneficiaries, enrollments, and hospitalizations, each identified by primary keys, and collapses repeated yearly records for each entity. Cleaning applies documented rules to resolve cross-year inconsistencies surfaced during normalization, such as assigning the modal value to conflicting sex or race codes, retaining the earliest reported birth year, and recoding irreconcilable values as Unknown. Finally, materialization produces the analysis-ready tables used in this study, including the mortality denominator, related denominator products, and hospitalization outcome datasets. At this stage, analysis-specific rules are applied, including truncating implausible ages, capping age at 115 for beneficiaries with no recorded death date, and imputing pre-2011 dual-eligibility status from the state buy-in variable. The full implementation is available in public repositories at \url{https://github.com/NSAPH-Data-Processing/}.

Lastly, in Appendix \ref{app:ssec:mobility} we explore summaries and graphical illustrations comparing residential and mobility-based exposures. Overall, we find that the average value of $\tau$ is 0.78 with most zip codes falling between 0.7 and 0.9, while the correlation between $\boldsymbol{W}$ and $\boldsymbol{G}$ is high. Mobility-based exposures are generally higher than residential ones, highlighting that individuals travel to areas of higher pollution. Given these findings, it is likely that mobility will not drastically change our findings, though could still provide material differences in effect estimates of interest. 

\subsection{Plausibility of identifying assumptions}
\label{ssec:Plausibility}

Our analysis rests on three key identifying assumptions and it is important to discuss their plausibility within the Medicare cohort. The first assumption is positivity and it is important to note that estimands such as $\omega(\boldsymbol{\Delta})$ that shift exposure from their observed values to a nearby value, necessarily rely on weaker positivity assumptions than standard estimands that fix exposure at a particular value for all units in the population being studied. Nonetheless, violations of positivity can occur and it is important to assess this possibility further. Additionally, this is the one identifying assumption that is empirically verifiable, though there is not a well-established approach for evaluating positivity with multivariate, continuous treatments such as those seen here. In Appendix \ref{app:ssec:positivity}, we propose an approach to assessing positivity that is based on values of the estimated generalized propensity score \citep{imai2004causal}. Overall, our analyses suggest that the positivity assumption is not violated for the exposure shifts being examined here. 

No unmeasured confounding is arguably the strongest assumption made in studies such as ours. We have collected a range of demographic and socioeconomic factors for each zip code to provide increased confidence in this assumption. Additionally, a recent analysis of the health effects of PM$_{2.5}$ in the Medicare cohort, which adjusted for additional covariates available from an auxiliary data set showed that effects remained similar whether these additional confounders were included \citep{makar2017estimating}. Nonetheless, this is still a strong assumption and there could be unmeasured variables that could bias our resulting estimates. Because of this possibility, in Section \ref{sec:Differencing} we develop a modified version of our approach that controls for all time-invariant confounding whether it is observed or not, by utilizing repeated data over time on the same zip codes. We find that our results remain similar even in this more robust analysis, leading to increased belief that our findings are not driven solely by unmeasured confounding. We additionally develop sensitivity analyses to the presence of unmeasured confounding, which we describe in Section \ref{sec:Differencing} and Appendix \ref{app:unmeasuredconfounding}. 

The final assumption is SUTVA, which we have already relaxed by allowing for interference based on where individuals travel. It could be violated, however, if our choice of $g(\cdot)$ is insufficient and the exposure mapping is misspecified. We argue that for estimating the health effects of long-term exposure to pollution, a weighted average of exposures across zip codes based on how much time is spent there is reasonable. When examining short-term exposure to air pollution, other functions, such as the maximum pollution exposure experienced during mobility might be more relevant as extreme exposures are more likely to drive short-term health effects. Additionally, we could extend the SUTVA assumption to allow for the potential outcomes to depend on both the weighted average of the neighbors exposures and other functions of the neighboring exposures. We believe, however, that doing so is not likely to increase the plausibility of the assumption and would make estimation more difficult by increasing the dimension of $\boldsymbol{G}$.

\subsection{The health effects of air pollution mixture}\label{sec:analysis_US}

We now estimate the causal effect of air pollution on mortality using the basis expansion and BART approaches from Section \ref{sec: Estimation} for each year separately. Results from the IPW approach show similar results and can be found in Appendix \ref{app:BinaryTreatment}. We run MCMC for 10,000 iterations, discarding the first 5,000 as a burn-in, and thinning every tenth sample. As a comparison, we also fit the same models with only $\bW$, which corresponds to ignoring mobility. To mitigate positivity violations common in multivariate exposure settings \citep{antonelli2024causal}, we assess the effect of a small shift in each pollutant of 0.05 standard deviations for each exposure.
\begin{figure}
    \centering
    \includegraphics[width=0.85\linewidth]{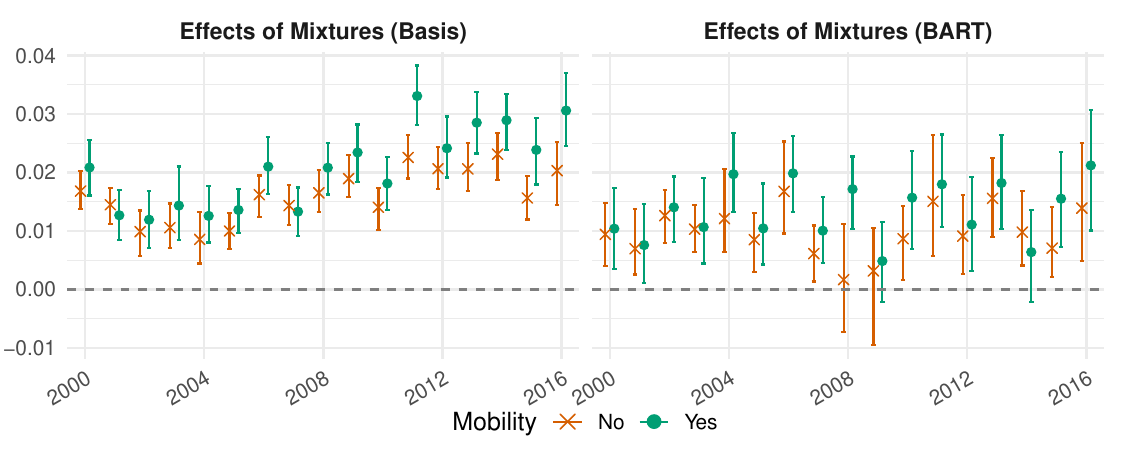}
    \caption{Point estimates and 95\% credible intervals for estimating the effect of increasing the air pollution mixture on mortality rates in the Medicare population.}
    \label{fig:analysis_US}
\end{figure}
Figure \ref{fig:analysis_US} shows point estimates and 95\% credible intervals for both model specifications. For the basis expansion model, whether we account for mobility or not, we found a positive and statistically significant estimate across all study years, with averages of 0.02 and 0.016 for the model with and without mobility, respectively. This implies about 20 additional deaths per 100,000 beneficiaries associated with a 0.05–standard–deviation increase in the PM$_{2.5}$ mixture. The BART model also yields significant, adverse effects of air pollution on mortality, though the estimated magnitudes are smaller, with averages of 0.013 and 0.01 for the models with and without mobility, respectively. The combination of these results provides evidence of a detrimental effect of increased air pollution exposure on mortality, which aligns with existing results in the literature. Interestingly, incorporating mobility leads to larger effect sizes, with average increases of 25\% and 30\% for the basis expansion and BART approaches, respectively. This, along with our results from Section \ref{sec:Bias}, suggests a bias towards the null when ignoring mobility. While the difference is not large, likely due to high correlations between $\boldsymbol{W}$ and $\boldsymbol{G}$, it is not negligible and suggests that accounting for mobility could be useful for increasing power to detect effects, or in studies with lower correlation between home and mobility-based exposures. Most existing studies ignore mobility and therefore our results suggest that the true health impacts of air pollution may have been underestimated in these existing studies. 

\subsection{Heterogeneous effects of air pollution mixture}
\label{ssec:MedicareHetero}

\begin{figure}
    \centering
    \includegraphics[width=0.88\linewidth]{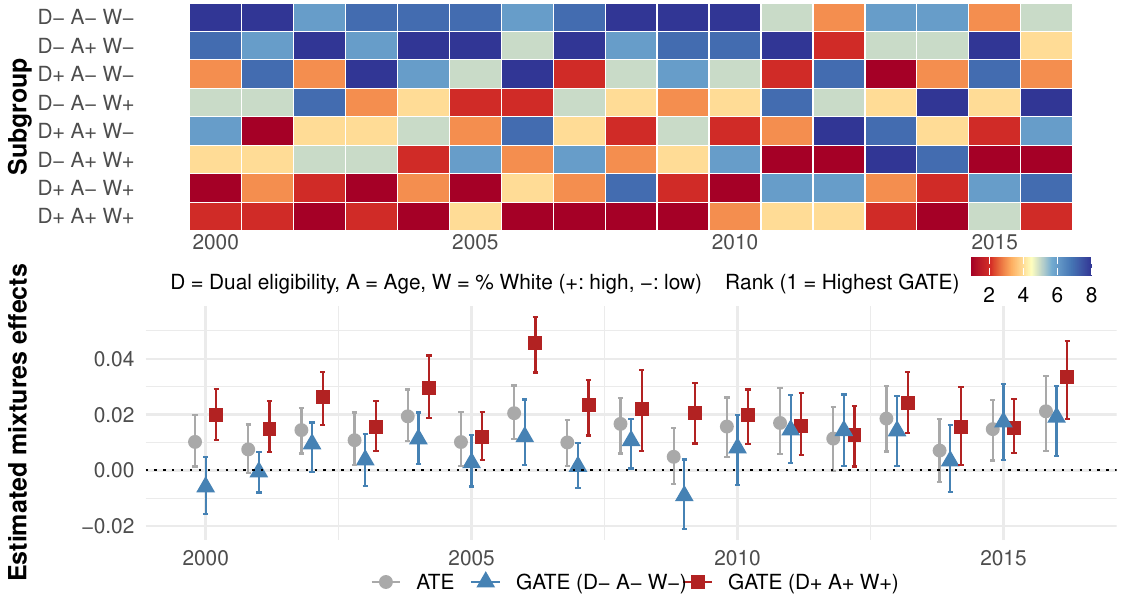}
    \caption{(Top) Yearly rank of estimated treatment effects by subgroup. (Bottom) Sample average treatment effect, treatment effects for groups with the highest rank and the lowest rank.}
    \label{fig:rank_gate}
\end{figure}
While we observe consistently harmful effects of increased air pollution mixtures, we further investigate whether these effects vary by zip code characteristics. We focus on the BART model, which flexibly captures complex treatment effect heterogeneity. For interpretability, we examine three key covariates commonly discussed in the literature \citep{bargagli2020causal,shin2025Treatmenta}: the rate of dual eligibility to Medicaid (a proxy for low socioeconomic status), mean age, and the proportion of White residents. Each year, each zip code is assigned to one of eight subgroups (two groups for each of the three covariates) based on whether its value is above or below the yearly median. We then calculate $\omega_s(\bDelta)$, as defined in Section \ref{sec: estimands}, for each subgroup. The upper panel in Figure \ref{fig:rank_gate} presents the yearly ranks of the group average treatment effects, with groups ordered by their mean rank across study years. Among the eight subgroups, zip codes with higher rates of dual eligibility to Medicaid, older populations, and higher proportions of White residents show the most adverse effects of increases in the air pollution mixture, whereas zip codes with the opposite characteristics---lower dual eligibility, younger populations, and fewer White residents---experience smaller effects. In the lower panel in Figure \ref{fig:rank_gate}, we compare the estimated effects for the overall highest- and lowest-ranked groups with the overall average treatment effect across all units. We find that the average effect for the highest-ranked group is significantly greater than that for the lowest group in several years, with a mean difference of 0.014. This corresponds to 14 additional deaths per 100,000 Medicare beneficiaries attributable to the air pollution increase in zip codes with higher rates of dual eligibility for Medicaid, older populations, and higher proportions of White residents, compared with zip codes with the opposite characteristics. Overall, our findings extend prior evidence of the pronounced health impacts of increased air pollution among older and low–socioeconomic–status populations \citep{bell2013Evidence} by incorporating components of \pmtpfs and accounting for population mobility. It is important to note, however, that our GATE estimand compares the impact of a common exposure \textit{shift}, though baseline exposure levels may differ across groups. Therefore each group's estimate corresponds to a slightly different estimand, though this still provides reasonably strong evidence that the effects of the air pollution mixture are indeed heterogeneous.

\subsection{Estimating the effect of each component}
\label{ssec:MedicareComponents}

\begin{figure}
    \centering
    \includegraphics[width=0.9\linewidth]{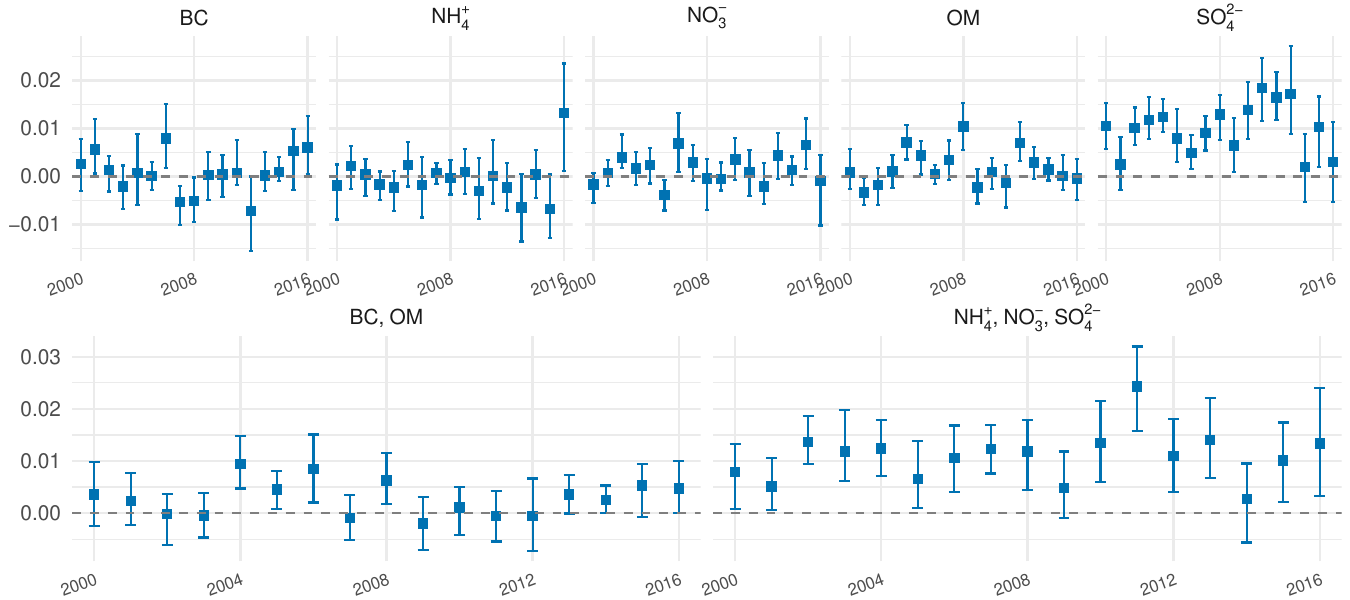}
    \caption{Estimated effects of each component (top) and a group of components (bottom)}
    \label{fig:component}
\end{figure}

To gain further insight into which pollutants drive the adverse mixture effect, we estimate the impact of shifting each \pmtpfs component separately while holding others fixed. We also examine the joint effects of two pollutant groups: (BC, OM) and (NH$_4^+$, NO$_3^-$, SO$_4^{2-}$) by shifting only the exposures within a group, while holding the other group constant. The first group represents primary carbonaceous pollutants emitted directly from combustion sources whereas the second group consists of secondary inorganic aerosols formed through atmospheric chemical reactions involving gaseous precursors. We again let each exposure be shifted by a 0.05 standard deviation increase to help avoid issues with positivity violations. These analyses allow us to identify which components are driving the adverse health effects estimated in Section \ref{sec:analysis_US}.

The results from the BART estimator are summarized in Figure \ref{fig:component}. For the individual-component analyses, BART shows relatively small to no effects for most components, but a consistently adverse effect of SO$_4^{2-}$. For the grouped analysis, BART indicates that the (NH$_4^+$, NO$_3^-$, SO$_4^{2-}$) group is the primary contributor to the adverse mixture effects seen in Section \ref{sec:analysis_US}, with some evidence of smaller effects for the (BC, OM) group. These results suggest that secondary inorganic aerosols, and SO$_4^{2-}$ in particular, are important contributors to the estimated adverse health effects of the \pmtpfs mixture.

\subsection{Control for time-invariant unmeasured confounders}
\label{sec:Differencing}

\begin{figure}
    \centering
    \includegraphics[width=0.8\linewidth]{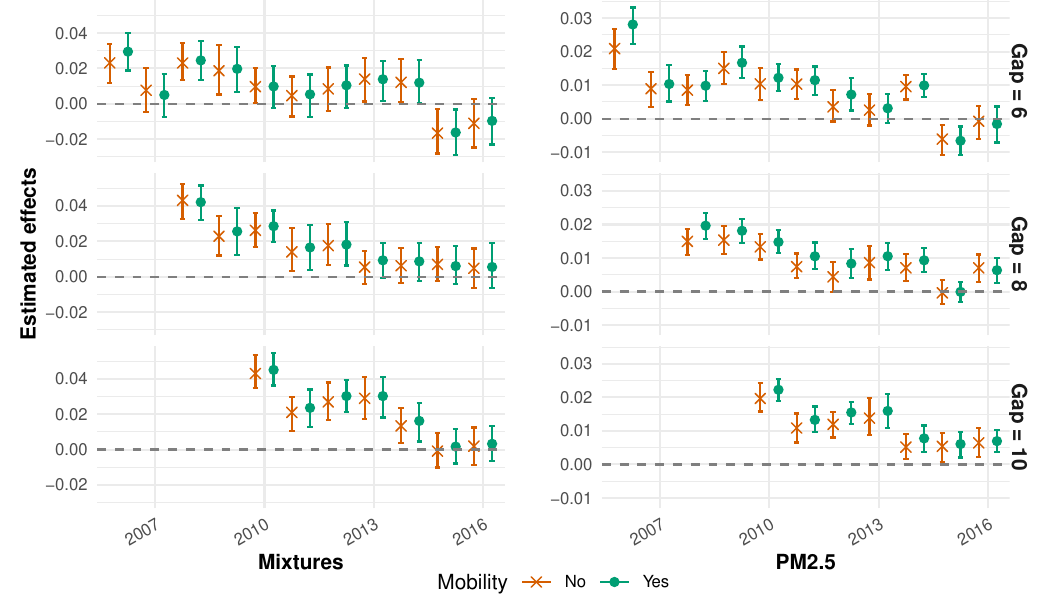}
    \caption{Effect estimates based on differencing across time with different gap lengths.}
    \label{fig:diff}
\end{figure}

Even though our yearly estimates adjusted for a comprehensive set of demographic, socioeconomic, and environmental covariates, there may still exist unmeasured factors correlated with both exposures and outcome that can bias the effect estimates. Here, to address time-invariant confounding, we extend a recently proposed approach for estimating treatment effects for continuous treatments with panel data that incorporates differencing across time to remove time-invariant confounding \citep{klosin2025Estimating}. Note that throughout we describe the approach in the context of our basis function model. Extending to BART is a useful avenue for future research, but would require novel computational techniques for implementation. Specifically, we assume the observed outcome follows
\begin{align*}
    Y_{it} &= a_i + \bX_{it} \boldsymbol{\theta} + \sum_{j=1}^q \Big\{ \tau_i \boldsymbol{\phi}_{wjt} \boldsymbol{\beta}_j + (1 - \tau_i) \boldsymbol{\phi}_{gjt} (\boldsymbol{\beta}_j + \boldsymbol{\zeta}_j) \Big\} + \epsilon_{it}.
\end{align*}
where a zip code level fixed effect $a_i$ captures unmeasured confounders that are fixed over time. In the presence of unmeasured confounders, estimates of the regression coefficients would be biased if we ignored $a_i$. To address this, we consider differencing between years $t$ and $t-l$:
{\small
\begin{align*}
    Y_{it}-Y_{i,t-l} 
    &= (\bx_{it}-\bx_{i(t-l)}) \boldsymbol{\theta} + \sum_{j=1}^q \Big\{ \tau_i (\boldsymbol{\phi}_{wjt} - \boldsymbol{\phi}_{wj(t-l)}) \boldsymbol{\beta}_j + (1 - \tau_i) (\boldsymbol{\phi}_{gjt}-\boldsymbol{\phi}_{gj(t-l)}) (\boldsymbol{\beta}_j + \boldsymbol{\zeta}_j) \Big\}
\end{align*}
}
This shows that we can model the differences in the outcomes between two time points using the difference in the covariates and in the exposure values. The coefficients of the model remain the same, but the differencing removes the time-invariant component $a_i$, thereby removing bias due to unmeasured zip code level confounders that do not vary over time. 

One key issue inherent to this approach in our setting is the length of time that we take differences over (denoted by $l$), and how long it takes for the effects of air pollution exposure to realize. If only the prior years exposure is expected to affect the outcome, then differencing across adjacent years (setting $l=1$) is sufficient. If, however, exposure effects take longer to realize and additional prior years of exposure can affect the outcome, then differencing across adjacent years will remove these effects and will only target the effect of the prior year's exposure. To avoid this issue, one can take differences over longer time periods in order to capture the effect of longer term exposure to air pollution. This is important in studies such as this one, because it is likely that the effects of pollution on mortality take longer than one year to realize. While we have used the prior years exposure to air pollution for $\boldsymbol{W}$, this effectively acts as a proxy for the exposure level in that zip code over prior years as well, since pollution levels are highly correlated across time. Because of this, if we take differences across adjacent years, we will be estimating the effect that a change in pollution in one year has on a change mortality in the following year, which will be very small or zero since it likely takes longer for long-term exposure to affect mortality. For this reason, we consider differencing across 6,8, and 10 year time gaps. Note that an additional approach to addressing this issue would be to explicitly include multiple prior years of exposure in the model as well, such as in distributed lag models, though incorporating this within our framework is beyond the scope of this work.

The results can be seen in Figure \ref{fig:diff}, where we see smaller effects for shorter time lags, but generally positive and significant effects for longer time lags. Note also that there is a decreasing trend in the magnitude of the exposure effects over time. This could be due to the effects of pollution truly decreasing over time, possibly due to improved public health or increasing awareness of the adverse health effects of pollution leading people to avoid exposure during elevated periods. Alternatively, this could be due to the presence of time-varying confounding, which this approach is not able to rule out.

Overall, these results significantly strengthen the evidence in support of an adverse causal effect of the air pollution mixture on mortality. Arguably the biggest assumption made by our prior analyses was the no unmeasured confounding assumption, and this analysis weakens that assumption substantially by allowing for arbitrary unmeasured confounders that are time-invariant. To further assess robustness to unmeasured confounding, we also conduct two additional sensitivity analyses. First, in Appendix \ref{app:unmeasuredconfounding}, we develop a sensitivity analysis framework to evaluate how strong an unmeasured confounder would need to be to materially alter the estimated effects. We find that our results are robust to unmeasured confounders that are as strongly associated with the exposure and outcome as the strongest of the observed covariates. As an additional sensitivity analysis, we fit a stacked outcome model with state-by-year fixed effects,
$$\E(Y_{it}\mid \boldsymbol W_{it},\boldsymbol G_{it},\boldsymbol X_{it})
=
\alpha_{s(i),t} + \bX_{it} \boldsymbol{\theta} + \sum_{j=1}^q \Big\{ \tau_i \boldsymbol{\phi}_{wjt} \boldsymbol{\beta}_j + (1 - \tau_i) \boldsymbol{\phi}_{gjt} (\boldsymbol{\beta}_j + \boldsymbol{\zeta}_j) \Big\},$$
where $\alpha_{s(i),t}$ allows the baseline mortality rate to vary across state-years, controlling for a broad class of state-level time-varying unmeasured confounders. This analysis yielded positive estimates for both the air-pollution mixture (0.0168 with mobility; 0.0147 without mobility) and PM$_{2.5}$ (0.0065 with mobility; 0.0054 without mobility), further supporting our main findings.

\section{Discussion}\label{sec:Discussion}
We introduced a novel approach to estimating the causal effects of multivariate exposures in the presence of spatial interference due to mobility of individuals across different regions. By incorporating cell phone mobility data, we obtain improved estimates of the impact of air pollution on mortality. We proposed policy-relevant causal estimands, and derived bias formulas for them when the no-interference assumption is wrongly assumed. Theoretical results and simulation studies suggest that the proposed model gives us a better assessment of the causal effects of air pollution, and that ignoring mobility will likely lead to under-estimation of the effects of air pollution. Lastly, we applied our framework to a national study of the health effects of air pollution mixtures on mortality among Medicare beneficiaries in the United States, and found a significant impact of air pollution on mortality. Additionally, incorporating mobility tended to increase the magnitude of these effect estimates, suggesting this could be useful for studies moving forward.

While this paper is the first to incorporate people's mobility into air pollution analysis using an interference framework, several limitations and future research directions remain. One drawback of our study is that the cell phone mobility data comes from the general United States population in 2019, whereas our inferences are on the Medicare population in the years 2000-2016. Because the Medicare cohort includes individuals over age 65, their mobility patterns likely differ from the general population. We derived robustness results protecting against misspecification of mobility, and our simulations and sensitivity analyses in the Appendix suggest that accounting for mobility with misspecified mobility patterns is still better than ignoring mobility altogether. We also note that while estimates incorporating mobility are slightly larger in general, the main findings from our study that show significant effects of the air pollution mixture hold whether mobility is accounted for or not, which provides assurance that the results seen are not driven solely by misspecified estimates of mobility patterns. Despite this, future research should aim to find mobility data that more closely aligns with the population being studied. A potential future area of research where mobility data could additionally provide benefits is the distinction between indoor and outdoor exposure to air pollution. Our exposure data consist of outdoor exposure to pollution, but mobility data could help infer the percentage of time individuals spend inside where exposure is decreased, which would likely provide more accurate estimates of the health effects of pollution. It is important to note, however, that while we focused on causal effects, our findings about the impact of mobility would also apply to general (non-causal) conditional associations of air pollutant health effects, and are therefore applicable to a wide range of environmental health studies. 

\bibliographystyle{apalike}

\bibliography{ref}

\appendix

\section{Illustration with linear models}
\label{sec:appendix_BiasLinear}

We first begin with a simple example with univariate exposures, and to simplify the calculations, we scale both $W$ and $G$ to have marginal variance 1 and correlation $\rho$. We assume that the potential outcomes are generated from the model given by
$$Y(w,g) = \tau w \beta_w + (1-\tau) g \beta_g + \epsilon$$
where $\tau$ is the proportion of time spent in the residential zip code. We explore two approaches first: 1) a linear model regressing the outcome against  $W$, and 2) a linear model regressing the outcome against $W^* = \tau W + (1 - \tau) G$, which can be seen as the true pollution exposure for a unit based on their time spent in different regions. We define the parameters from these two models as
\begin{align*}
    (\tilde{\beta}_0, \tilde{\beta}_w) &= \underset{\beta_0, \beta}{\operatorname{argmin}} \ \E [(Y - \beta_0 - W \beta)^2]   \\
    (\tilde{\beta}_0^*, \tilde{\beta}_w^*) &= \underset{\beta_0, \beta}{\operatorname{argmin}} \ \E [(Y - \beta_0 - W^* \beta)^2].
\end{align*}
Our focus will be on the parameters $\tilde{\beta}_w$ and $\tilde{\beta}_w^*$. The first of these parameters corresponds to a na\"ive analysis that ignores mobility and relates pollution to health only through the exposure at the residential region, which is common practice in environmental health epidemiology. The second parameter of interest, $\tilde{\beta}_w^*$ captures the impact of using the true weighted average exposure based on movement throughout the day. In Appendix \ref{sec:BiasProof}, we show these are given by the following expressions:
\begin{align*}
    \tilde{\beta}_w &= \tau \beta_w + \rho (1 - \tau) \beta_g \\
    \tilde{\beta}_w^* &= \frac{\tau^2 \beta_w  + \rho \tau (1 - \tau) \beta_w  + \rho \tau (1 - \tau) \beta_g + (1 - \tau)^2 \beta_g}{\tau^2 + 2 \tau (1 - \tau) \rho + (1 - \tau)^2 }.
\end{align*}
Now suppose that we are interested in estimating the effect of increasing pollution everywhere by one unit, which amounts to increasing both $W$ and $G$ by 1. The true causal effect in this case would be $\tau \beta_w + (1 - \tau)\beta_g$. Unsurprisingly, the na\"ive approach, $\tilde{\beta}_w$, is biased except in the extreme case where $\rho = 1$. Importantly, If $\beta_w$ and $\beta_g$ have the same sign, which is reasonable in environmental applications, the na\"ive approach will underestimate the effect of pollution. 

Moving to the second parameter, we show in Appendix \ref{sec:BiasProof} that $\tilde{\beta}_w^*$ is equal to the true effect if any one of three conditions hold: 1) $\rho = 1$, 2) $\tau \in  \{0, 0.5, 1\}$, or 3) $\beta_w = \beta_g$. The first two conditions are unlikely to hold in practice as we don't expect $W$ and $G$ to be perfectly correlated, or for people to spend exactly 0\%, 50\%, or 100\% of the time in their residential region. The final condition implies that pollution is equally harmful in residential and neighboring areas, which is reasonable in some settings, but can fail to hold in others. One way in which these effects could differ is if people spend more time outside while in neighboring areas, whereas they spend more time inside when in their residential area. Indoor and outdoor pollution can vary substantially, which would lead to differences in the impact of these two pollution sources. This shows the importance of separating the effects of $W$ and $G$ as it provides unbiased estimates of health effects regardless of the value of $(\tau, \rho, \beta_w, \beta_g)$. 

\section{Derivation of bias results}\label{sec:BiasProof}

Here we provide justification for the bias calculations seen in Section \ref{sec: linear}. Recall that we investigate the following two traditional approaches:
\begin{align*}
    (\tilde{\beta}_0, \tilde{\beta}_w) &= \underset{\beta_0, \beta}{\operatorname{argmin}} \ \E [(Y - \beta_0 - W \beta)^2]   \\
    (\tilde{\beta}_0^*, \tilde{\beta}_w^*) &= \underset{\beta_0, \beta}{\operatorname{argmin}} \ \E [(Y - \beta_0 - W^* \beta)^2].
\end{align*}
We write
\begin{align*}
    \E [(Y - \beta_0 - W \beta)^2] &= \E [(\tau W \beta_w + (1-\tau) G \beta_g + \epsilon - \beta_0 - W \beta)^2]\\
    &=\E [W^2\beta^2 - 2 W \beta \{\tau W \beta_x + (1-\tau) G \beta_g - \beta_0 + \epsilon \} \\
    &\hspace{0.5 in} + \{\tau W \beta_x + (1-\tau) G \beta_g - \beta_0 + \epsilon \}^2]\\
    &=\beta^2 - 2\beta \E[\tau W^2 \beta_x + (1-\tau) WG \beta_g - W\beta_0 + W\epsilon ]  + C\\
    &=\beta^2 - 2\beta \E[\tau W^2 \beta_x + (1-\tau) WG \beta_g]  + C
\end{align*}
for some constant $C$. Therefore,
\begin{align*}
    \tilde{\beta}_w = \E[\tau W^2 \beta_x + (1-\tau) WG \beta_g] = \tau\beta_x + (1-\tau)\rho\beta_g.
\end{align*}
Similarly, since
\begin{align*}
    \E [(Y - \beta_0 - W^* \beta)^2] &= \E [[\tau W \beta_w + (1-\tau) G \beta_g + \epsilon - \beta_0 - \{\tau W + (1-\tau) G\} \beta]^2]\\
    &= \beta_0^2 - 2\beta_0 \E [\tau W \beta_w + (1-\tau) G \beta_g + \epsilon - \{\tau W + (1-\tau) G\} \beta] + C\\
    &= \beta_0^2 + C,
\end{align*}
$\tilde{\beta}_0^*=0$, and given $\tilde{\beta}_0=\tilde{\beta}_0^*=0$, we can write
\begin{align*}
    \E [(Y - \beta_0 - W^* \beta)^2] &= \E [[\tau W \beta_w + (1-\tau) G \beta_g + \epsilon - \{\tau W + (1-\tau) G\} \beta]^2]\\
    &=\E [\{\tau W + (1-\tau) G\}^2] \beta^2 \\
    &\hspace{0.5 in}- 2 \E[\{\tau W + (1-\tau) G\}\{\tau W \beta_w + (1-\tau) G \beta_g + \epsilon \}] \beta + C.
\end{align*}
Therefore, we have
\begin{align*}
    \tilde{\beta}_w^* &= \frac{\E[\{\tau W + (1-\tau) G\}\{\tau W \beta_w + (1-\tau) G \beta_g + \epsilon \}]}{\E[\{\tau W + (1-\tau) G\}^2]}\\
    &= \frac{\tau^2\beta_w + \tau(1-\tau)\rho\beta_g + \tau(1-\tau)\rho\beta_w + (1-\tau)^2\beta_g}{\tau^2 + 2\tau(1-\tau)\rho + (1-\tau)^2}.
\end{align*}

Our next goal is to show under what conditions $\tilde{\beta}_w^*$ equals the true causal effect of $\tau \beta_w + (1 - \tau) \beta_g$. To do so, we can take the difference and set it equal to zero:
\begin{align*}
    & \frac{\tau^2 \beta_w  + \rho \tau (1 - \tau) \beta_w  + \rho \tau (1 - \tau) \beta_g + (1 - \tau)^2 \beta_g}{\tau^2 + 2 \tau (1 - \tau) + \rho (1 - \tau)^2 } - (\tau \beta_w + (1 - \tau)\beta_g) = 0 \\
    \Rightarrow \ & \tau^2 \beta_w  + \rho \tau (1 - \tau) \beta_w  + \rho \tau (1 - \tau) \beta_g + (1 - \tau)^2 \beta_g \\
    & - (\tau \beta_w + (1 - \tau)\beta_g)(\tau^2 + 2 \tau (1 - \tau) + \rho (1 - \tau)^2)  = 0  \\
    \Rightarrow \ & \beta_w [ 2 (\rho - 1) \tau^3  - 3(\rho - 1) \tau^2  + (\rho - 1) \tau ] + \beta_g [ -2 (\rho - 1) \tau^3  + 3(\rho - 1) \tau^2  - (\rho - 1) \tau ] = 0 \\
    \Rightarrow \ & \tau (\tau - 1) (2\tau - 1) (\beta_w - \beta_g) (\rho - 1)   = 0.
\end{align*}

This final expression shows that the parameter takes the desired value when any of the following three conditions hold: 1) $\rho = 1$, 2) $\tau \in  \{0, 0.5, 1\}$, or 3) $\beta_w = \beta_g$.

\section{Misspecification of weights proof}\label{sec: miss_bias_proof}
In this section, we prove the bias results presented in Section \ref{sec:MissWeights} where we misspecify the travel weights $\tau$. Let $\tau^*_i$ denote the true proportion of time spent in the home region and we incorrectly assume that it is $\tau_i$. Recall that the true outcome function is given by
$$Y_i = \beta_w^* \tau_i^* W_i + \beta_g^* (1 - \tau_i^*) G_i + \epsilon_i.$$
Consider we fit the model with the misspecified $\tau$.
Define $\widetilde\bW=(\tau_1W_1,...,\tau_nW_n)'$, $\widetilde\bG=((1-\tau_1)G_1,...,(1-\tau_n)G_n)'$, $\widetilde\bW^*=(\tau_1^*W_1,...,\tau_n^*W_n)'$, and $\widetilde\bG^*=((1-\tau_1^*)G_1,...,(1-\tau_n^*)G_n)'$. The expectation of the OLS estimator of the regression coefficients, $\widehat\bbeta=(\widehat\beta_w, \widehat\beta_g)'$, is given by
\begin{align*}
    \E(\widehat\bbeta|\widetilde\bW,\widetilde\bG)
    &= \begin{pmatrix}
        \widetilde\bW'\widetilde\bW & \widetilde\bW'\widetilde\bG\\
        \widetilde\bW'\widetilde\bG & \widetilde\bG'\widetilde\bG
    \end{pmatrix}^{-1} \begin{pmatrix}
        \widetilde\bW'\widetilde\bW^* & \widetilde\bW'\widetilde\bG^*\\
        \widetilde\bG'\widetilde\bW^* & \widetilde\bG'\widetilde\bG^*
    \end{pmatrix}\bbeta^*\\
    &= \frac{1}{K} \begin{pmatrix}
        \widetilde\bG'\widetilde\bG & -\widetilde\bW'\widetilde\bG\\
        -\widetilde\bW'\widetilde\bG & \widetilde\bW'\widetilde\bW
    \end{pmatrix} \begin{pmatrix}
        \widetilde\bW'\widetilde\bW^* & \widetilde\bW'\widetilde\bG^*\\
        \widetilde\bG'\widetilde\bW^* & \widetilde\bG'\widetilde\bG^*
    \end{pmatrix}\bbeta^*\\
    &= \frac{1}{K} \begin{pmatrix}
        \widetilde\bG'\widetilde\bG\widetilde\bW'\widetilde\bW^* -\widetilde\bW'\widetilde\bG\widetilde\bG'\widetilde\bW^* & \widetilde\bG'\widetilde\bG\widetilde\bW'\widetilde\bG^*-\widetilde\bW'\widetilde\bG\widetilde\bG'\widetilde\bG^*\\
        -\widetilde\bW'\widetilde\bG\widetilde\bW'\widetilde\bW^* + \widetilde\bW'\widetilde\bW\widetilde\bG'\widetilde\bW^* & -\widetilde\bW'\widetilde\bG\widetilde\bW'\widetilde\bG^* + \widetilde\bW'\widetilde\bW\widetilde\bG'\widetilde\bG^*
    \end{pmatrix} \bbeta^*
\end{align*}
where $K=\norm{\widetilde\bW}^2\norm{\widetilde\bG}^2 - (\widetilde\bW'\widetilde\bG)^2$ and $\norm{\cdot}$ denotes the $l^2$ norm. Therefore, the expectation of the estimator of $\omega(\bDelta)$ is
\begin{align*}
    \E(\widehat{\omega})&= \E(\widehat{\beta}_w)\widebar{\tau} + \E(\widehat{\beta}_g) (1 - \widebar{\tau})\\
    &= \frac{1}{K}\Big[ \big\{(\widetilde\bG'\widetilde\bG\widetilde\bW'\widetilde\bW^* - \widetilde\bW'\widetilde\bG\widetilde\bG'\widetilde\bW^*) \beta_w^* + (\widetilde\bG'\widetilde\bG\widetilde\bW'\widetilde\bG^*-\widetilde\bW'\widetilde\bG\widetilde\bG'\widetilde\bG^*) \beta_g^* \big\} \widebar{\tau} \\
    &+ \big\{(-\widetilde\bW'\widetilde\bG\widetilde\bW'\widetilde\bW^* + \widetilde\bW'\widetilde\bW\widetilde\bG'\widetilde\bW^*) \beta_w^* + (-\widetilde\bW'\widetilde\bG\widetilde\bW'\widetilde\bG^* + \widetilde\bW'\widetilde\bW\widetilde\bG'\widetilde\bG^*) \beta_g^* \big\} (1-\widebar{\tau}) \Big]
\end{align*}
where the expectation is taken with respect to the residuals in the outcome model. Hence, the bias is given by
\begin{align*}
    \E(\widehat{\omega}-\omega) &= \beta_w^*\bigg\{ \dfrac{\widetilde\bG'\widetilde\bG\widetilde\bW'\widetilde\bW^* - \widetilde\bW'\widetilde\bG\widetilde\bG'\widetilde\bW^*}{\widetilde\bW'\widetilde\bW\widetilde\bG'\widetilde\bG - \widetilde\bW'\widetilde\bG\widetilde\bW'\widetilde\bG} \widebar{\tau} + \dfrac{\widetilde\bW'\widetilde\bW\widetilde\bG'\widetilde\bW^* -\widetilde\bW'\widetilde\bG\widetilde\bW'\widetilde\bW^*}{\widetilde\bW'\widetilde\bW\widetilde\bG'\widetilde\bG - \widetilde\bW'\widetilde\bG\widetilde\bW'\widetilde\bG} (1-\widebar{\tau})  - \widebar{\tau^*}\bigg\}\\
        & + \beta_g^* \bigg\{ \dfrac{\widetilde\bG'\widetilde\bG\widetilde\bW'\widetilde\bG^*-\widetilde\bW'\widetilde\bG\widetilde\bG'\widetilde\bG^*}{\widetilde\bW'\widetilde\bW\widetilde\bG'\widetilde\bG - \widetilde\bW'\widetilde\bG\widetilde\bW'\widetilde\bG} \widebar{\tau} + \dfrac{ \widetilde\bW'\widetilde\bW\widetilde\bG'\widetilde\bG^*-\widetilde\bW'\widetilde\bG\widetilde\bW'\widetilde\bG^*}{\widetilde\bW'\widetilde\bW\widetilde\bG'\widetilde\bG - \widetilde\bW'\widetilde\bG\widetilde\bW'\widetilde\bG}(1-\widebar{\tau})  - (1 -\widebar{\tau^*})\bigg\}.
\end{align*}

\subsection{\texorpdfstring{$\tau^*=c\tau$} {tauStar=ctau} case}
If we assume that $\tau^*$ and $\tau$ are independent of the exposures, and that the exposures are standardized, then we can take the limit of the bias
expression above as $n \rightarrow\infty$ to obtain

\begin{align}\label{eq: Bias_asym}
\begin{split}
    & \beta_w^*\Bigg\{ \dfrac{\Big( \E[(1 - \tau)^2] \E(\tau \tau^*) - \rho^2 \E[\tau (1 - \tau)] \E[\tau^* (1 - \tau)] \Big) \E(\tau)}{\E(\tau^2) \E[(1 - \tau)^2] - \rho^2 \E^2[\tau (1 - \tau)]}\\
    &\qquad+\dfrac{\Big( \rho \E[\tau^2] \E[\tau^* (1 - \tau)] - \rho \E[\tau (1 - \tau)] \E[\tau \tau^*] \Big) (1 -\E(\tau))}{\E(\tau^2) \E[(1 - \tau)^2] - \rho^2 \E^2[\tau (1 - \tau)]}  - \E(\tau^*)\Bigg\} +\\
    &\beta_g^* \Bigg\{ \dfrac{\Big( \rho \E[(1 - \tau)^2] \E[\tau (1 - \tau^*)] - \rho \E[\tau (1 - \tau)] \E[(1 - \tau) (1 - \tau^*)] \Big) \E(\tau)}{\E(\tau^2) \E[(1 - \tau)^2] - \rho^2 \E^2[\tau (1 - \tau)]}  \\
    &\qquad+\dfrac{\Big( \E(\tau^2) \E[(1 - \tau)(1 - \tau^*)] - \rho^2 \E[\tau (1 - \tau)] \E[\tau (1 - \tau^*)] \Big) (1 -\E(\tau))}{\E(\tau^2) \E[(1 - \tau)^2] - \rho^2 \E^2[\tau (1 - \tau)]}  - (1 - \E(\tau^*))\Bigg\}.
\end{split}
\end{align}
There are two terms inside the brackets of the $\beta_w^*$ term and two for the $\beta_g^*$ term. Let's assume that $\tau = \tau^*/c$ for some constant $c>0$ which is equivalent to $\tau^* = c \tau$. Now let's examine the first term:
\begin{align*}
    & \dfrac{\Big( \E[(1 - \tau)^2] \E(\tau \tau^*) - \rho^2 \E[\tau (1 - \tau)] \E[\tau^* (1 - \tau)] \Big) \E(\tau) }{\E(\tau^2) \E[(1 - \tau)^2] - \rho^2 \E^2[\tau (1 - \tau)]} \\
    = & \dfrac{\Big( c \E[(1 - \tau)^2] \E(\tau^2) - c\rho^2 \E[\tau (1 - \tau)] \E[\tau (1 - \tau)] \Big) \E(\tau) }{\E(\tau^2) \E[(1 - \tau)^2] - \rho^2 \E^2[\tau (1 - \tau)]} \\
    = & c \E(\tau) = \E(\tau^*)
\end{align*}

Similarly, we can look at the second term
\begin{align*}
   & \dfrac{\Big( \rho \E[\tau^2] \E[\tau^* (1 - \tau)] - \rho \E[\tau (1 - \tau)] \E[\tau \tau^*] \Big) (1 -\E(\tau))}{\E(\tau^2) \E[(1 - \tau)^2] - \rho^2 \E^2[\tau (1 - \tau)]} \\
   = & \dfrac{\Big( c \rho \E[\tau^2] \E[\tau (1 - \tau)] - c \rho \E[\tau (1 - \tau)] \E[\tau^2] \Big) (1 -\E(\tau))}{\E(\tau^2) \E[(1 - \tau)^2] - \rho^2 \E^2[\tau (1 - \tau)]} \\
   = & 0
\end{align*}
The combination of these results shows that the term involving $\beta_w^*$ is exactly zero in this scenario, since we also subtract $\E(\tau^*)$ from these terms, when the truth is a constant multiple of the estimated $\tau$. A trickier issue is to deal with the terms in the $\beta_g^*$ component of the bias, which don't disappear so easily.
The denominator is the same in both of the terms, so we focus on the numerator of the two terms, starting with the first one:
\begin{align*}
    & \Big( \rho \E[(1 - \tau)^2] \E[\tau (1 - \tau^*)] - \rho \E[\tau (1 - \tau)] \E[(1 - \tau) (1 - \tau^*)] \Big) \E(\tau) \\ = & \ \Big( \rho \E[1 - 2\tau + \tau^2] \E[\tau - c\tau^2)] - \rho \E[\tau - \tau^2] \E[1 - (1 + c)\tau + c \tau^2] \Big) \E(\tau) \\
    = & \ \rho (1 - c) [\E(\tau^2) - \E^2(\tau)] \E(\tau) \\
    = & \ \rho (1 - c) \text{Var}(\tau) \E(\tau)
\end{align*}

Now we can do the same thing but for the numerator of the second term:
\begin{align*}
    & \Big( \E(\tau^2) \E[(1 - \tau)(1 - \tau^*)] - \rho^2 \E[\tau (1 - \tau)] \E[\tau (1 - \tau^*)] \Big) (1 -\E(\tau)) \\
    = & \ \Big( \E(\tau^2) \E[(1 - (1 + c)\tau + c\tau^2)] - \rho^2 \E[\tau - \tau^2] \E[\tau - c\tau^2)] \Big) (1 -\E(\tau)) \\
    = & \ \Big( \E(\tau^2) - \rho^2 \E^2(\tau) + (1 + c) (\rho^2 - 1) \E(\tau) \E(\tau^2) + c (1 - \rho^2) \E^2(\tau^2) \Big) (1 -\E(\tau))
\end{align*}

Lastly, we can expand out the denominator of the term involving $\beta_g^*$:
\begin{align*}
    & \E(\tau^2) \E[(1 - \tau)^2] - \rho^2 \E^2[\tau (1 - \tau)] \\
    = & \ \E(\tau^2) - 2\E(\tau^2) \E(\tau) + \E^2(\tau^2) - \rho^2 \E^2(\tau) + 2 \rho^2 \E(\tau) \E(\tau^2) - \rho^2 \E^2(\tau^2) \\
    = & \E(\tau^2) - 2 (1 - \rho^2) \E(\tau) \E(\tau^2) + (1 - \rho^2) \E^2(\tau^2) - \rho^2 \E^2(\tau)
\end{align*}

We need the ratio of the numerator and the denominator to equal $1 - \E(\tau)^* = 1 - c\E(\tau)$. Alternatively, we need the numerator to be equal to the denominator times $1 - c\E(\tau)$. We aim to show that the numerator is similar to the denominator times $1 - c\E(\tau)$. First, we can look at the numerator, which is the sum of the two numerator terms above:
\begin{align*}
    & \ \rho(1 - c) \E(\tau) \E(\tau^2) - \rho (1 - c) \E^3(\tau) + \E(\tau^2) - \rho^2 \E^2(\tau) + (1 + c) (\rho^2 - 1) \E(\tau) \E(\tau^2) \\
    + & \ c (1 - \rho^2) \E^2(\tau^2) - \E(\tau) \E(\tau^2) + \rho^2 \E^3(\tau) - (1 + c)(\rho^2 - 1) \E^2(\tau) \E(\tau^2) - c (1 - \rho^2)\E(\tau) \E^2(\tau^2) \\
    = & \ \E(\tau^2) + c (1 - \rho^2) \E^2(\tau^2) - \rho^2 \E^2(\tau) + [\rho (1 - c) - (1 + c) (1 - \rho^2) - 1] \E(\tau)\E(\tau^2) \\
    - & \ c (1 - \rho^2)\E(\tau) \E^2(\tau^2) + [\rho^2 - \rho(1 - c)] \E^3(\tau) - (1 + c)(\rho^2 - 1) \E^2(\tau) \E(\tau^2)
\end{align*}

Now we can do the same thing with the denominator. We will take the denominator and multiply by $1 - c E(\tau)$. Doing so gives us the following:
\begin{align*}
    & \ \E(\tau^2) - 2 (1 - \rho^2) \E(\tau) \E(\tau^2) + (1 - \rho^2) \E^2(\tau^2) - \rho^2 \E^2(\tau) \\
    - & \ c \E(\tau) \E(\tau^2) + 2c(1 - \rho^2)\E^2(\tau) \E(\tau^2) - c (1 - \rho^2) \E(\tau) \E^2(\tau^2) + c \rho^2 \E^3 (\tau) \\
    = & \ \E(\tau^2) + (1 - \rho^2) \E^2(\tau^2) - \rho^2 \E^2(\tau) + [-c - 2(1 - \rho^2)] \E(\tau) \E(\tau^2) \\
    - & \ c(1 - \rho^2) \E(\tau) \E^2(\tau^2) + c \rho^2 \E^3 (\tau) + 2c(1 - \rho^2)\E^2(\tau) \E(\tau^2)
\end{align*}

We want the numerator and denominator to be similar. A good sanity check is to see that if $c=1$, they become the same and the bias becomes zero. Clearly, however, the bias is not zero as these terms are not equivalent when $c \neq 1$. When $\rho=1$ this bias also goes to zero, but we want a result that holds in general. We can look at the numerator and denominator to see where they differ and better understand the degree of bias. We can write the ratio of the numerator and the denominator times $1 - c E(\tau)$ as $\frac{D + B}{D}$, and we want this ratio to be 1. The magnitude of the bias is therefore given by $\beta_g^* B / D$. Let's investigate $B$ to learn more about this bias. It is fairly straightforward to show that we have
\begin{align*}
    B &= -(1 - c) (1 - \rho^2) E^2(\tau^2) + (1 - c)(\rho - \rho^2) E(\tau) E(\tau^2) \\
    & + (1 - c) (1 - \rho^2) E^2(\tau) E(\tau^2) - (1 - c) (\rho - \rho^2) E^3(\tau) \\
    &= (1 - c) (1 - \rho) \text{Var}(\tau) \Big[ \rho E(\tau) - (1 + \rho) E(\tau^2) \Big]
\end{align*}

And to see the bias, we can take the ratio of this with the denominator in its original form:

\begin{align*}
    \text{Bias} &= \frac{\beta_g^* (1 - c) (1 - \rho) \text{Var}(\tau) \Big[ \rho E(\tau) - (1 + \rho) E(\tau^2) \Big]}{\Big(E(\tau^2) E[(1 - \tau)^2] - \rho^2 E^2[\tau (1 - \tau)] \Big)}
\end{align*}

\begin{figure}
    \centering
    \includegraphics[width=4.5 in]{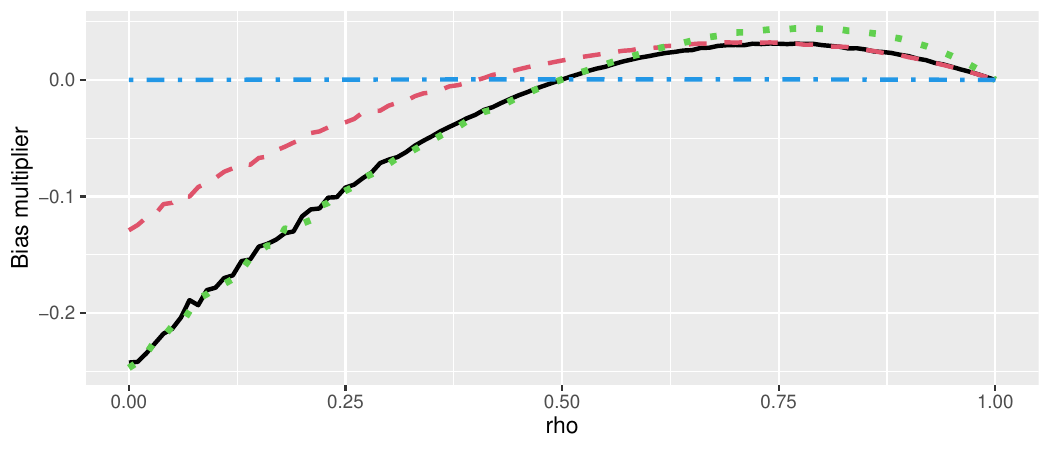}
    \caption{Bias by $\rho$ when $c=0.5$. Each line represents a different distribution of $\tau$. All distributions are beta distributions with parameters $a_{\tau}$ and $b_{\tau}$. Black corresponds to $a_{\tau} = 0.006$ and $b_{\tau} = 0.006$, red corresponds to $a_{\tau} = 1$ and $b_{\tau} = 1$, green corresponds to $a_{\tau} = 148$ and $b_{\tau} = 1$, and blue corresponds to $a_{\tau} = 1$ and $b_{\tau} = 148$.}
    \label{fig: bias_multiplier}
\end{figure}

Figure \ref{fig: bias_multiplier} illustrates the behavior of the bias expression above when $\beta_g^*=1$. We find that when the correlation between home and neighborhood exposures is greater than 0.4, which is reasonable in many environmental studies as we have seen in Section \ref{sec:mobility}, the magnitude of the bias can be bounded (in absolute value) by 0.05 times $\beta_g^*$ across many distributions for $\tau$.

\subsection{\texorpdfstring{$\tau$}{tau} with measurement error}
From Equation \ref{eq: Bias_asym}, if we assume that $\tau^*_i = \tau_i + \eta_i \in [0,1]$ for all $i=1,...,n$ where $\eta_i$'s are independent and identically distributed measurement errors with mean 0 and also independent of $\tau_i$, the bias expression becomes 
\begin{align*}
    & \beta_w^*\Bigg\{ \dfrac{\Big( E[(1 - \tau)^2] E(\tau (\tau+\eta)) - \rho^2 E[\tau (1 - \tau)] E[(\tau+\eta) (1 - \tau)] \Big) E(\tau)}{E(\tau^2) E[(1 - \tau)^2] - \rho^2 E^2[\tau (1 - \tau)]}\\
    &\qquad+\dfrac{\Big( \rho E[\tau^2] E[\tau+\eta (1 - \tau)] - \rho E[\tau (1 - \tau)] E[\tau (\tau+\eta)] \Big) (1 -E(\tau))}{E(\tau^2) E[(1 - \tau)^2] - \rho^2 E^2[\tau (1 - \tau)]}  - E(\tau+\eta)\Bigg\} +\\
    &\beta_g^* \Bigg\{ \dfrac{\Big( \rho E[(1 - \tau)^2] E[\tau (1 - \tau+\eta)] - \rho E[\tau (1 - \tau)] E[(1 - \tau) (1 - \tau-\eta)] \Big) E(\tau)}{E(\tau^2) E[(1 - \tau)^2] - \rho^2 E^2[\tau (1 - \tau)]}  \\
    &\qquad+\dfrac{\Big( E(\tau^2) E[(1 - \tau)(1 - \tau+\eta)] - \rho^2 E[\tau (1 - \tau)] E[\tau (1 - \tau-\eta)] \Big) (1 -E(\tau))}{E(\tau^2) E[(1 - \tau)^2] - \rho^2 E^2[\tau (1 - \tau)]}  - (1 - E(\tau+\eta))\Bigg\}\\
    &=\beta_w^*\Bigg\{ \dfrac{\Big( E[(1 - \tau)^2] E(\tau^2) - \rho^2 E[\tau (1 - \tau)] E[\tau (1 - \tau)] \Big) E(\tau)}{E(\tau^2) E[(1 - \tau)^2] - \rho^2 E^2[\tau (1 - \tau)]}\\
    &\qquad+\dfrac{\Big( \rho E[\tau^2] E[\tau (1 - \tau)] - \rho E[\tau (1 - \tau)] E[\tau^2] \Big) (1 -E(\tau))}{E(\tau^2) E[(1 - \tau)^2] - \rho^2 E^2[\tau (1 - \tau)]}  - E(\tau)\Bigg\} +\\
    &\beta_g^* \Bigg\{ \dfrac{\Big( \rho E[(1 - \tau)^2] E[\tau (1 - \tau)] - \rho E[\tau (1 - \tau)] E[(1 - \tau)^2] \Big) E(\tau)}{E(\tau^2) E[(1 - \tau)^2] - \rho^2 E^2[\tau (1 - \tau)]}  \\
    &\qquad+\dfrac{\Big( E(\tau^2) E[(1 - \tau)^2] - \rho^2 E[\tau (1 - \tau)] E[\tau (1 - \tau)] \Big) (1 -E(\tau))}{E(\tau^2) E[(1 - \tau)^2] - \rho^2 E^2[\tau (1 - \tau)]}  - (1 - E(\tau))\Bigg\}\\
    &=\beta_w^*\{E(\tau)+0-E(\tau)\} +\beta_g^*\{0+(1-E(\tau))-(1-E(\tau))\}\\
    &=0.
\end{align*}
Therefore, we get an unbiased estimate with such measurement errors.

On the other hand, if we consider $\tau_i=\tau^*_i+\eta_i$ for all $i$, where $\eta_i$ represents independent and identically distributed measurement errors with mean 0 and a finite second moment, the bias expression in Equation \ref{eq: Bias_asym} becomes significantly more complex. However, to provide further insight, let us imagine a specific scenario where the correlation between the home exposure $W$ and neighborhood exposure $G$ is assumed to be zero ($\rho=0$), which is something of a worst case scenario for the bias. Then, the bias expression is given by
\begin{align*}
    & \beta_w^*\Bigg\{ \dfrac{E[(1 - \tau)^2] E(\tau \tau^*)}{E(\tau^2) E[(1 - \tau)^2]}\E(\tau) - \widebar{\tau^*}\Bigg\} +\beta_g^* \Bigg\{ \dfrac{E(\tau^2) E[(1 - \tau)(1 - \tau^*)]}{E(\tau^2) E[(1 - \tau)^2]}(1 -\widebar{\tau})  - (1 - \widebar{\tau^*})\Bigg\}\\
    &=\beta_w^*\Bigg\{ \dfrac{E[(1 - \tau^*-\eta)^2] E((\tau*+\eta) \tau^*)}{E((\tau^*+\eta)^2) E[(1 - \tau^*-\eta)^2]}E(\tau^*+\eta) - E(\tau^*)\Bigg\} \\
    &\qquad+\beta_g^* \Bigg\{ \dfrac{E((\tau^*+\eta)^2) E[(1 - \tau^*-\eta)(1 - \tau^*)]}{E((\tau^*+\eta)^2) E[(1 - \tau^*-\eta)^2]}(1 -E(\tau^*+\eta)) - (1 - E(\tau^*))\Bigg\}\\
    &=\beta_w^*\Bigg\{ \dfrac{E[(1 - \tau^*-\eta)^2] E({\tau^*}^2)}{E((\tau^*+\eta)^2) E[(1 - \tau^*-\eta)^2]}E(\tau^*) - E(\tau^*)\Bigg\} \\
    &\qquad+\beta_g^* \Bigg\{ \dfrac{E((\tau^*+\eta)^2) E[(1 - \tau^*)^2]}{E((\tau^*+\eta)^2) E[(1 - \tau^*-\eta)^2]}E(1 -\tau^*) - E(1 - \tau^*)\Bigg\}\\
    &\equiv \beta_w^*\{\xi_wE(\tau^*) - E(\tau^*)\} + \beta_g^*\{\xi_gE(1 -\tau^*) - E(1 - \tau^*)\}.
\end{align*}

\begin{figure}
    \centering
    \includegraphics[width=6.5 in]{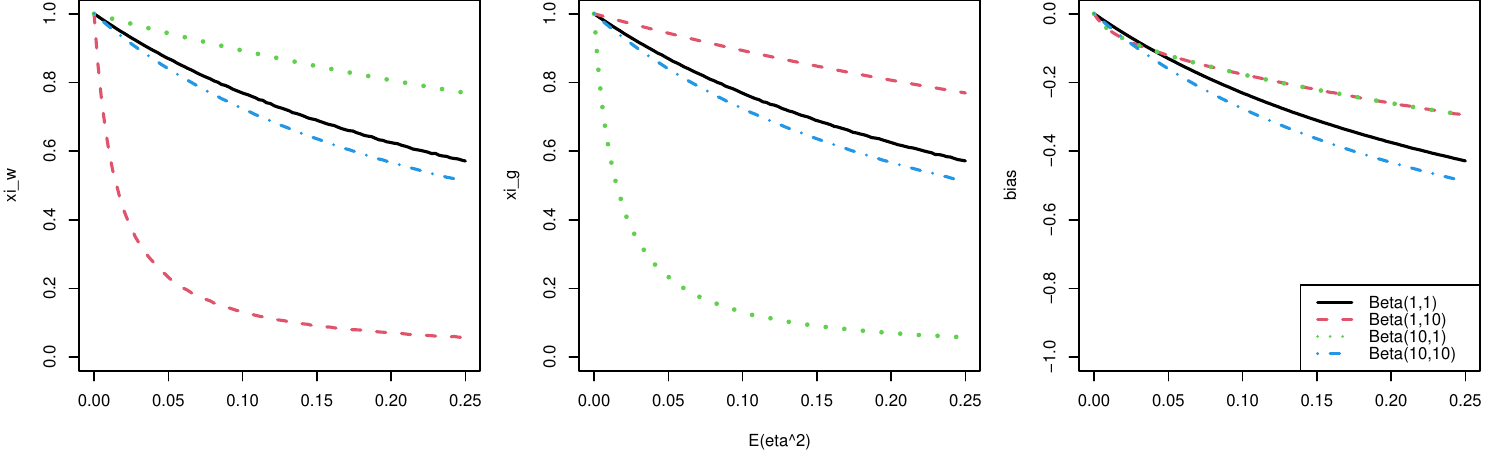}
    \caption{Illustration of behavior of $\xi_w$ and $\xi_g$ as a function of the variance of measurement error $\E(\eta^2)$ under different distributions for $\tau^*$. Asymptotic bias is calculated under $\beta_w^*=\beta_g^*=1$.}
    \label{fig:xi_plot}
\end{figure}

Next, we claim that $\xi_w$ and $\xi_g$ are continuous decreasing functions of $\E(\eta^2)$ with positive second derivatives and the limits of 0 as $\E(\eta^2)\rightarrow\infty$.
\begin{proof}
We can rewrite $\xi_w$ as
\begin{align*}
    \xi_w &= \dfrac{\{\E[(1-\tau^*)^2] + \E(\eta^2)\}\E({\tau^*}^2)}{\E({\tau^*}^2)\E[(1-\tau^*)^2] + \{\E({\tau^*}^2) + \E[(1-\tau^*)^2])\}\E(\eta^2) + \E^2(\eta^2)}\\
    &= \dfrac{\E({\tau^*}^2)\E[(1-\tau^*)^2] + \E({\tau^*}^2)\E(\eta^2)}{\E({\tau^*}^2)\E[(1-\tau^*)^2] + \{\E({\tau^*}^2) + \E[(1-\tau^*)^2])\}\E(\eta^2) + \E^2(\eta^2)}.
\end{align*}
Therefore, $\lim_{\E(\eta^2)\rightarrow 0}\xi_w=1$ and $\lim_{\E(\eta^2)\rightarrow\infty}\xi_w=0$.\

Next, the partial derivative with respect to $\E(\eta^2)$ is
\begin{align*}
    &\dfrac{\partial\xi_w}{\partial\E(\eta^2)}\\
    &= \bigg(\E({\tau^*}^2)\big[\E({\tau^*}^2)\E[(1-\tau^*)^2] + \{\E({\tau^*}^2) + \E[(1-\tau^*)^2])\}\E(\eta^2) + \E^2(\eta^2)\big]\\
    &\qquad - \{\E({\tau^*}^2)\E[(1-\tau^*)^2] + \E({\tau^*}^2)\E(\eta^2)\}\{\E({\tau^*}^2) + \E[(1-\tau^*)^2] + 2\E(\eta^2)\}\bigg)/Den\\
    &=\bigg(\E^2(\eta^2)\{\E({\tau^*}^2) -2\E({\tau^*}^2)\} + \\
    &\E(\eta^2)\{\E^2({\tau^*}^2) + \E({\tau^*}^2)\E[(1-\tau^*)^2] - 2\E({\tau^*}^2)\E[(1-\tau^*)^2]-\E^2({\tau^*}^2) - \E({\tau^*}^2)\E[(1-\tau^*)^2]\}+\\
    &\E^2({\tau^*}^2)\E[(1-\tau^*)^2] - \{\E^2({\tau^*}^2)\E[(1-\tau^*)^2] - \E({\tau^*}^2)\E^2[(1-\tau^*)^2]\}\bigg)/Den\\
    &= \dfrac{-\E({\tau^*}^2)\E^2(\eta^2) - 2\E({\tau^*}^2)\E[(1-\tau^*)^2]\E(\eta^2) - \E({\tau^*}^2)\E^2[(1-\tau^*)^2]}{\big(\E^2(\eta^2) + \{\E({\tau^*}^2) + \E[(1-\tau^*)^2])\}\E(\eta^2) + \E({\tau^*}^2)\E[(1-\tau^*)^2]\big)^2} \leq 0
\end{align*}
for $\E(\eta^2)\geq 0$ where $Den=\big(\E^2(\eta^2) + \{\E({\tau^*}^2) + \E[(1-\tau^*)^2])\}\E(\eta^2) + \E({\tau^*}^2)\E[(1-\tau^*)^2]\big)^2$. Therefore, $\xi_w$ is a decreasing function of $\E(\eta^2)\geq 0$. Also,
$$\dfrac{\partial\xi_w}{\partial\E(\eta^2)}\eval_{\E(\eta^2)=0}=-\dfrac{1}{\E({\tau^*}^2)} = -\dfrac{1}{\E^2(\tau^*) + \var(\tau^*)}$$
which together with the following concave property implies that $\xi_w$ is robust to the measurement error if the mean of $\tau^*$ is close to 0 and has small variability. We now show that $\xi_w$ is a concave function of $\E(\eta^2)\geq 0$.
\scriptsize{
\begin{align*}
    &\dfrac{\partial^2\xi_w}{\partial\E^2(\eta^2)}\\
    &= \bigg(-2\Big\{\E({\tau^*}^2)\E(\eta^2) + \E({\tau^*}^2)\E[(1-\tau^*)^2]\Big\}\Big\{\E^2(\eta^2) + \{\E({\tau^*}^2) + \E[(1-\tau^*)^2])\}\E(\eta^2) + \E({\tau^*}^2)\E[(1-\tau^*)^2]\Big\}^2 + \\
    & 2\Big\{ \E({\tau^*}^2)\E^2(\eta^2) + 2\E({\tau^*}^2)\E[(1-\tau^*)^2]\E(\eta^2) + \E({\tau^*}^2)\E^2[(1-\tau^*)^2]\Big\}\Big\{\E^2(\eta^2) + \{\E({\tau^*}^2) + \E[(1-\tau^*)^2])\}\E(\eta^2) + \E({\tau^*}^2)\E[(1-\tau^*)^2]\Big\}\\
    &\Big\{\{\E({\tau^*}^2) + \E[(1-\tau^*)^2])\} + 2\E(\eta^2)\Big\}\bigg)/Den\\
    &= \bigg(-\Big\{\E({\tau^*}^2)\E(\eta^2) + \E({\tau^*}^2)\E[(1-\tau^*)^2]\Big\}\Big\{\E^2(\eta^2) + \{\E({\tau^*}^2) + \E[(1-\tau^*)^2])\}\E(\eta^2) + \E({\tau^*}^2)\E[(1-\tau^*)^2]\Big\}\\
    & + \Big\{\E({\tau^*}^2)\E(\eta^2) + 2\E({\tau^*}^2)\E[(1-\tau^*)^2]\E(\eta^2) + \E({\tau^*}^2)\E^2[(1-\tau^*)^2]\Big\}\Big\{\{\E({\tau^*}^2) + \E[(1-\tau^*)^2])\} + 2\E(\eta^2)\Big\}\bigg)/Den'\\
    &= (-AB+A'B')/Den' \geq 0
\end{align*}
}
\normalsize
One can see that $0\leq A\leq A'$, $0\leq B\leq B'$, and $Den'>0$, which suggests that $\xi_w$ is concave which concludes the proof. An analogous proof for $\xi_g$ is straightforward.
\end{proof}
Figure \ref{fig:xi_plot} illustrates the behavior of $\xi_w$, $\xi_g$, and the bias when $\beta_w^*=\beta_g^*=1$. As an example, assuming a uniform distribution of $\eta$ between -0.25 and 0.25, which suggests a significant amount of measurement error, the expected value of $\eta^2$ is approximately 0.02. This leads to a magnitude of bias that remains below 0.08 under a number of distributions for $\tau^*$.

\section{Prior specification and posterior computation for the basis-expansion model}\label{sec: full conditional derivation}

In this section, we provide details for posterior computation under the basis-expansion model described in Section \ref{sec: Estimation}. We first state the prior specification and then derive the full conditional distributions used in the Gibbs sampler.

Recall that we consider a structured semiparametric model that directly encodes the mobility decomposition through a basis expansion. Specifically, we model the conditional mean as
\begin{align*}
    \E (Y | \bW, \bG, \bX; \tau) 
    &= \bx_i \boldsymbol{\theta} 
    + \tau_i \sum_{j=1}^q\sum_{m=1}^M \beta_{jm}\phi_m(w_{ij}) 
    + (1 - \tau_i) \sum_{j=1}^q\sum_{m=1}^M (\beta_{jm} + \zeta_{jm})\phi_m(g_{ij}) \\
    &=  \bx_i \boldsymbol{\theta} 
    + \sum_{j=1}^q \Big\{ 
    \tau_i \boldsymbol{\phi}_{w,ij}^\top \boldsymbol{\beta}_j 
    + (1 - \tau_i) \boldsymbol{\phi}_{g,ij}^\top(\boldsymbol{\beta}_j + \boldsymbol{\zeta}_j) 
    \Big\},
\end{align*}
where $\{\phi_m(\cdot)\}_{m=1}^M$ are orthogonal basis functions, 
$\boldsymbol{\phi}_{w,ij}=(\phi_1(w_{ij}),\ldots,\phi_M(w_{ij}))^\top$, and 
$\boldsymbol{\phi}_{g,ij}$ is defined analogously. The common coefficient vector 
$\boldsymbol{\beta}$ contributes to the effects of both the home and mobility-based exposures, while $\boldsymbol{\zeta}$ captures deviations between these two effects. Thus, the model encodes the idea that home and mobility-based exposures may have similar, but not necessarily identical, effects.

To encourage shrinkage across these two functions, we adopt the following horseshoe prior distribution \citep{carvalho2010horseshoe}:
\begin{align}\label{eq: horseshoe}
\begin{split}
        &\bbeta_j |\sigma_\beta^2 \sim MVN (\bo, \sigma^2\sigma_\beta^2 \bi), \qquad \sigma_\beta^2\sim IG(1,1)\\
    &\bzeta_j|\lambda_j^2, \nu^2 \sim MVN(\bo, \sigma^2\lambda_j^2 \nu^2 \bi), \qquad \lambda_j \sim C^{+}(1), \qquad \nu\sim C^{+}(1)\\
    &\sigma^2 \sim IG(1,1)
\end{split}
\end{align}
where $\sigma^2$ is the residual variance of $Y$, $\bzeta_j=(\zeta_{j1},\cdots,\zeta_{jM})'$, $MVN(\cdot,\cdot)$ is the multivariate normal distribution, $IG(\cdot,\cdot)$ is the Inverse-Gamma distribution, and $C^+(k)$ denotes the Half-Cauchy distribution with the density function $f(z)=1/\{\pi k (1+z^2/k^2)\}$. Although the Half-Cauchy distribution is not conditionally conjugate for the normal distribution, \cite{makalic2015simple} proposed an alternative way to construct the equivalent model by introducing an auxiliary variable with a double Inverse-Gamma structure. It uses the fact that if there are random variables such that $V^2|W \sim IG(1/2,1/W)$ and $W\sim IG(1/2,1/k^2)$, then $V\sim C^+(k)$. Thus, we define the following hierarchy which is equivalent to the one described in Equation \ref{eq: horseshoe}:
\begin{align*}
    &\bbeta_j |\sigma^2,\sigma_\beta^2 \sim MVN (\bo, \sigma^2\sigma_\beta^2 \bi), \qquad \sigma_\beta^2\sim IG(1,1)\\
    &\bzeta_j|\sigma^2, \lambda_j^2, \nu^2 \sim MVN(\bo, \sigma^2\lambda_j^2 \nu^2 \bi)\\
    &\lambda_j^2|r_j \sim IG(1/2, 1/r_j), \qquad r_j \sim IG(1/2, 1) \\
    &\nu^2|s \sim IG(1/2, 1/s), \qquad s\sim IG(1/2, 1) \\
    &\sigma^2 \sim IG(1,1)
\end{align*}

Now all priors above lead to conditionally conjugate posterior updates, and therefore Gibbs sampling from their full conditional posterior distributions is straightforward. We can write $\bY \sim MVN((\bphi_{w} + \bphi_{g})\bbeta + \bphi_{g}\bzeta, \sigma^2I)$ where $\bphi_{w}$ and $\bphi_{g}$ are $n\times qM$ matrices with the $i^{th}$ row of $\tau_i\bphi_{wi}$ and $(1-\tau_i)\bphi_{gi}$, respectively. Then, the full conditional posterior distributions are given by:
\begin{align*}
    &\bbeta|\cdot \sim MVN (\widetilde\bSigma_\beta(\bphi_{w} + \bphi_{g})'(\by - \bphi_{g}\bzeta)/\sigma^2, \widetilde\bSigma_\beta), \quad \sigma_\beta^2|\cdot \sim IG(1+qM/2, 1+ \Vert\bbeta\Vert_2^2/(2\sigma^2)\\
    &\bzeta|\cdot \sim MVN(\widetilde\bSigma_\zeta \{\bphi_{g}'(\by - (\bphi_{w} + \bphi_{g})\bbeta)/\sigma^2 \}, \widetilde\bSigma_\zeta)\\
    &\lambda_j^2|\cdot \sim IG((1+M)/2, 1/r_j + \Vert\bzeta_j\Vert_2^2/(2\sigma^2\nu^2)), \quad r_j|\cdot \sim IG(1,1+1/\lambda_j^2) \\
    &\nu^2|\cdot \sim IG((1+qM)/2, 1/s + \bzeta'\bLambda^{-1}\bzeta/(2\sigma^2)), \quad s|\cdot\sim IG(1,1+1/\nu^2) \\
    &\sigma^2|\cdot \sim IG(1 + qM + n/2, 1 + \{\Vert \by-(\bphi_{w} + \bphi_{g})\bbeta - \bphi_{g}\bzeta \Vert_2^2 + \Vert\bbeta\Vert_2^2/\sigma_\beta^2 + \bzeta'\bLambda_*^{-1}\bzeta\}/2)
\end{align*}
where $\widetilde\bSigma_\beta=\{(\bphi_{w} + \bphi_{g})'(\bphi_{w} + \bphi_{g})/\sigma^2 + 1/(\sigma^2\sigma_\beta^2)\}^{-1}$. $\widetilde\bSigma_\zeta = \sigma^2\{\bphi_{g}'\bphi_{g} + \bLambda_*^{-1}\}^{-1}$ where $\bLambda_*=\nu^2\bLambda$ and $\bLambda=\diag(\lambda_1^2,\cdots,\lambda_1^2,\lambda_2^2,\cdots,\lambda_q^2)$, a $qM$-dimensional diagonal matrix with the elements being $M$ replicates of each $\lambda_j$. Now we go into more detail of how to derive each full conditional distribution. The full conditional density of $\bbeta$ given all the other parameters and data is
\begin{align*}
    \pi(\bbeta|\cdot) &\propto p(\by|\cdot) \times \pi(\bbeta|\sigma^2, \sigma_\beta^2)\\
    &\propto \exp\bigg(-\dfrac{1}{2\sigma^2} \Vert \by-(\bphi_{w} + \bphi_{g})\bbeta - \bphi_{g}\bzeta \Vert_2^2  - \dfrac{1}{2\sigma^2\sigma_\beta^2}\Vert\bbeta\Vert_2^2 \bigg)\\
    &\propto \exp\bigg(-\dfrac{1}{2}\Big[ \bbeta'\{(\bphi_{w} + \bphi_{g})'(\bphi_{w} + \bphi_{g})/\sigma^2 + 1/(\sigma^2\sigma_\beta^2)\}\bbeta -2\bbeta'(\bphi_{w} + \bphi_{g})'(\by - \bphi_{g}\bzeta)/\sigma^2 \Big] \bigg)
\end{align*}
which is a kernel of the multivariate normal distribution with the covariance matrix $\widetilde\bSigma_\beta = \{(\bphi_{w} + \bphi_{g})'(\bphi_{w} + \bphi_{g})/\sigma^2 + 1/(\sigma^2\sigma_\beta^2)\}^{-1}$ and the mean vector $\widetilde\mu_\beta = \widetilde\bSigma_\beta(\bphi_{w} + \bphi_{g})'(\by - \bphi_{g}\bzeta)/\sigma^2$. Since $\sigma_\beta^2$ affects the likelihood only through $\bbeta$, the full conditional density of $\sigma_\beta^2$ is
\begin{align*}
    \pi(\sigma_\beta^2|\cdot) &\propto  \pi(\bbeta|\sigma^2, \sigma_\beta^2) \times \pi(\sigma_\beta^2)\\
    &\propto \dfrac{1}{(\sigma_\beta^2)^{qM/2}}\exp\bigg(- \frac{\Vert\bbeta\Vert_2^2}{2\sigma^2\sigma_\beta^2} \bigg)\times \frac{1}{(\sigma_\beta^2)^{1+1}}\exp\bigg(- \frac{1}{\sigma_\beta^2}\bigg)\\
    &\propto \frac{1}{(\sigma_\beta^2)^{1+1+qM/2}}\exp\bigg(- \frac{1+\Vert\bbeta\Vert_2^2/(2\sigma^2)}{\sigma_\beta^2}\bigg).
\end{align*}
Thus, it follows $IG(1+qM/2, 1+ \Vert\bbeta\Vert_2^2/(2\sigma^2))$. For the full conditional distribution of $\bzeta=(\bzeta_1',\cdots,\bzeta_q')'$, we define $\bLambda=\diag(\lambda_1,\cdots,\lambda_1,\lambda_2,\cdots,\lambda_q)$, a $qM$-dimensional diagonal matrix with the elements of $M$ replicates of each $\lambda_j$, and $\bLambda_*=\nu^2\bLambda$. Given the independence priors for $\bzeta_j$'s, we observe
\begin{align*}
    \pi(\bzeta|\cdot) &\propto p(\by|\cdot) \times \prod_{j=1}^q\pi(\bzeta_j|\sigma^2, \lambda_j^2, \nu^2)\\
    &\propto \exp\bigg(-\dfrac{1}{2\sigma^2} \Vert \by-(\bphi_{w} + \bphi_{g})\bbeta - \bphi_{g}\bzeta \Vert_2^2  - \dfrac{1}{2\sigma^2}\bzeta'\bLambda_*^{-1}\bzeta \bigg)\\
    &\propto \exp\bigg(-\dfrac{1}{2}\Big[\bzeta'\Big\{\dfrac{\bphi_{g}'\bphi_{g} + \bLambda_*^{-1}}{\sigma^2} \Big\}\bzeta -2\bzeta'\{\bphi_{g}'(\by - (\bphi_{w} + \bphi_{g})\bbeta)/\sigma^2 \} \Big] \bigg),
\end{align*}
which is again a kernel of the multivariate normal distribution with the covariance matrix $\widetilde\bSigma_\zeta = \sigma^2\{\bphi_{g}'\bphi_{g} + \bLambda_*^{-1}\}^{-1}$ and the mean vector $\widetilde\mu_\zeta = \widetilde\bSigma_\zeta \{\bphi_{g}'(\by - (\bphi_{w} + \bphi_{g})\bbeta)/\sigma^2 \}$. The conditional posterior of the local and global variance parameters are conjugate as
\begin{align*}
    \pi(\lambda_j^2|\cdot)&\propto \pi(\bzeta_j|\sigma^2,\lambda_j^2, \nu^2) \times \pi(\lambda_j^2|r_j)\\
    &\propto \dfrac{1}{(\lambda_j^2)^{M/2}}\exp\bigg(-\dfrac{\Vert\bzeta_j\Vert_2^2}{2\sigma^2\lambda_j^2\nu^2}\bigg) \dfrac{1}{(\lambda_j^2)^{1/2+1}}\exp\bigg(-\dfrac{1/r_j}{\lambda_j^2}\bigg)\\
    &\propto \dfrac{1}{(\lambda_j^2)^{(1+M)/2 + 1}}\exp\bigg(-\dfrac{1/r_j + \Vert\bzeta_j\Vert_2^2/(2\sigma^2\nu^2)}{\lambda_j^2}\bigg)\\
    \pi(\nu^2|\cdot)&\propto \prod_{j=1}^q\pi(\bzeta_j|\sigma^2,\lambda_j^2, \nu^2) \times \pi(\nu^2|s)\\
    &\propto \dfrac{1}{(\nu^2)^{qM/2}}\exp\bigg(-\dfrac{\bzeta'\bLambda^{-1}\bzeta}{2\sigma^2\nu^2}\bigg) \dfrac{1}{(\nu^2)^{1/2+1}}\exp\bigg(-\dfrac{1/s}{\nu^2}\bigg)\\
    &\propto \dfrac{1}{(\nu^2)^{(1+qM)/2+1}}\exp\bigg(-\dfrac{1/s+\bzeta'\bLambda^{-1}\bzeta/(2\sigma^2)}{\nu^2}\bigg)
\end{align*}
and therefore, $\lambda_j^2|\cdot\sim IG((1+M)/2, 1/r_j + \Vert\bzeta_j\Vert_2^2/(2\sigma^2\nu^2))$ for all $j=1,...,q$ and $\nu^2|\cdot\sim IG((1+qM)/2, 1/s + \bzeta'\bLambda^{-1}\bzeta/(2\sigma^2))$. Further, the updates for the auxiliary variables are
\begin{align*}
    \pi(r_j|\cdot)&\propto \pi(\lambda_j^2|r_j) \times \pi(r_j)\\
    &\propto \dfrac{1}{r_j^{1/2}}\exp\bigg(-\dfrac{1/r_j}{\lambda_j^2}\bigg)\dfrac{1}{(r_j)^{1/2+1}}\exp\bigg(-\dfrac{1}{r_j}\bigg)\\
    &\propto\dfrac{1}{(r_j)^{1+1}}\exp\bigg(-\dfrac{1+1/\lambda_j^2}{r_j}\bigg)\\
    \pi(s|\cdot)&\propto \pi(\nu^2|s)\times\pi(s)\\
    &\propto \dfrac{1}{s^{1/2}}\exp\bigg(-\dfrac{1/s}{\nu^2}\bigg)\dfrac{1}{(s)^{1/2+1}}\exp\bigg(-\dfrac{1}{s}\bigg)\\
    &\propto\dfrac{1}{(s)^{1+1}}\exp\bigg(-\dfrac{1+1/\nu^2}{s}\bigg)
\end{align*}
which suggests that $r_j|\cdot\sim IG(1,1+1/\lambda_j^2)$ and $s|\cdot\sim IG(1,1+1/\nu^2)$. Finally, the full conditional posterior of the residual variance $\sigma^2$ is
\begin{align*}
    \pi(\sigma^2|\cdot) &\propto p(\by|\cdot) \times \pi(\bbeta|\sigma^2,\sigma_\beta^2) \times \prod_{j=1}^q\pi(\bzeta_j|\sigma^2, \lambda_j^2, \nu^2) \times \pi(\sigma^2) \\
    &\propto \dfrac{1}{(\sigma^2)^{n/2}}\exp\bigg(-\dfrac{1}{2\sigma^2} \Vert \by-(\bphi_{w} + \bphi_{g})\bbeta - \bphi_{g}\bzeta \Vert_2^2\bigg) \times \dfrac{1}{(\sigma^2)^{qM/2}}\exp\bigg(- \frac{\Vert\bbeta\Vert_2^2}{2\sigma^2\sigma_\beta^2} \bigg) \\
    &\qquad \times \dfrac{1}{(\sigma^2)^{qM/2}}\exp\bigg(-\dfrac{\bzeta'\bLambda_*^{-1}\bzeta}{2\sigma^2}\bigg) \times \dfrac{1}{(\sigma^2)^{1+1}}  \exp(-1 / \sigma^2)\\
    &\propto \dfrac{1}{(\sigma^2)^{1+1+qM+n/2}}\exp\bigg(-\dfrac{1 + \{\Vert \by-(\bphi_{w} + \bphi_{g})\bbeta - \bphi_{g}\bzeta \Vert_2^2 + \Vert\bbeta\Vert_2^2/\sigma_\beta^2 + \bzeta'\bLambda_*^{-1}\bzeta\}/2}{\sigma^2} \bigg)
\end{align*}
which is a kernel of $IG(1 + qM + n/2, 1 + \{\Vert \by-(\bphi_{w} + \bphi_{g})\bbeta - \bphi_{g}\bzeta \Vert_2^2 + \Vert\bbeta\Vert_2^2/\sigma_\beta^2 + \bzeta'\bLambda_*^{-1}\bzeta\}/2)$.

\section{Simulation Studies}\label{sec:Simulation}
In our simulation studies, we generate data in the following manner:
\begin{align*}
    &\bW_i \iidsim MVN(\bo_{q}, \Sigma_{q\times q}) \mathtext{for $i=1,...,n$ where} \Sigma_{jk}=\begin{cases}
    1 & j=k\\
    0.7^{|j-k|} & j\neq k
    \end{cases}
    .
\end{align*}
Throughout we set $q=5$, and $n=1000$. Given that correlation between pollutants is frequently observed in environmental studies, we focus on the case where pollutants are highly correlated. Next, we generate mobility weights $\tau_i$ from a $\nbeta(30,10)$ distribution, so that people travel outside of their home area for 75\% of the day on average. Further, we assume that people in each region travel to 10 different areas, i.e. $\sum_{j} 1(\alpha_{ij} \neq 0) = 10$, each having similar pollution levels to the home region in order to make home and neighborhood exposures positively correlated. This was done to match what we found empirically in Section \ref{sec:mobility}. Mathematically, for the $i^{th}$ region, we sample 10 indices from $(1,..., i-1,i+1,...n)$ with probabilities of $(p_{i,1},...,p_{i,i-1},p_{i,i+1},...,p_{i,n})$ where $p_{i,j}\propto \exp\big(-0.4\Vert \widetilde{\bW}_i - \widetilde{\bW}_j\Vert_2\big)$. This leads to an average correlation between home and neighborhood exposures of 0.7. After selecting where to travel, for each $j\in\mathcal{I}_i$ where $\mathcal{I}_i$ is an index set of travel regions of the $i^{th}$ region, a travel proportion $\alpha_{ij}$ is determined by its selection probability scaled by the sum of selection probabilities of the 10 regions, $p_{i,j}/\sum_{k \in\mathcal{I}_i} p_{i,k}$. Given $\tau$, $\balpha$, $\bW$, and $\bG$, we let the outcome be generated by
\begin{align*}
    Y_i&= \tau_i\bphi_{wi}\bbeta + (1-\tau_i)\bphi_{gi}\bgamma + \epsilon_i\\
    &= \tau_i\bphi_{wi}\bbeta + (1-\tau_i)\bphi_{gi}(\bbeta+\bzeta) + \epsilon_i
\end{align*}
where $\epsilon_i\iidsim N(0,\sigma^2)$, and $\bphi_{wi}$ and $\bphi_{gi}$ are $3^{rd}$-degree orthogonalized polynomial basis expansions of $\bW_i$ and $\bG_i$, respectively. We set 
$$\bbeta = (.5,.5,0,0,.3,.3,0,0,.7,.7,0,...,0)'$$
so the first 4 exposures have up to cubic effects on the outcome. We assume that the difference between the effects of home and neighborhood exposures follows $\bzeta\sim MVN(0, \sigma^2_\zeta\Vec{I})$. We expect the home and neighborhood exposures to have similar, but not necessarily the same, effects on the outcome, and therefore, we simulate three different scenarios: 1)  $\sigma_\zeta=0$ (\textit{No difference}), 2) $\sigma_\zeta=0.15$ (\textit{Small difference}), and 3) $\sigma_\zeta=0.3$ (\textit{Moderate difference}).

\begin{figure}
    \centering
    \includegraphics[width=6.5 in]{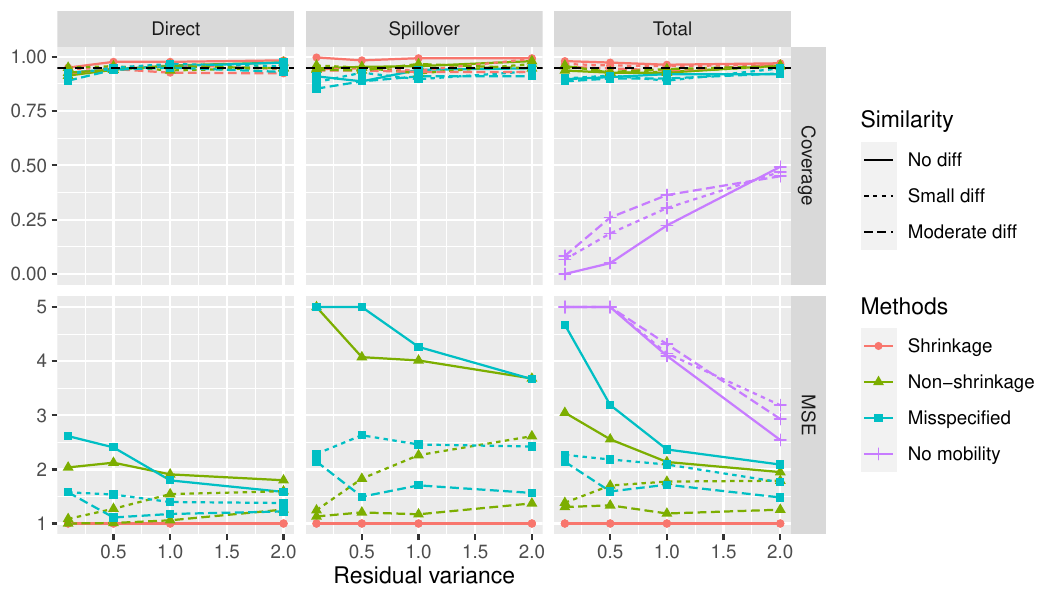}
    \caption{MSE and 95\% interval coverage across all simulation scenarios for each of the four estimators. Note that the MSE has been capped at 5 to improve visualization. }
    \label{fig:Sim_plot}
\end{figure}

Our goal is to understand the sample average causal effect of a 0.5 increment of each exposure in all areas, which corresponds to estimating $\omega(\bDelta)$ with $\bDelta_1=\cdots=\bDelta_n=(.5,...,.5)'$. For all approaches except for the naive approach, we also estimate $\omega_{dir}(\bDelta)$ and $\omega_{sp}(\bDelta)$, which can be defined by
\begin{align*}
    \omega(\bDelta) &= \frac{1}{n} \sum_{i=1}^n \bigg\{ Y_i(\bw_i + \bDelta_{wi}, \bg_i + \bDelta_{gi}) - Y_i(\bw_i, \bg_i) \bigg\} \\
    &= \frac{1}{n} \sum_{i=1}^n \big\{ Y_i(\bw_i + \bDelta_{wi}, \bg_i + \bDelta_{gi}) - Y_i(\bw_i + \bDelta_{wi}, \bg_i) \big\} \\
    &\hspace{0.2 in}+ \frac{1}{n} \sum_{i=1}^n \big\{ Y_i(\bw_i + \bDelta_{wi}, \bg_i) - Y_i(\bw_i, \bg_i) \big\} \\
    &\equiv \omega_{sp}(\bDelta) + \omega_{dir}(\bDelta).
\end{align*}
The results of the simulation are presented in Figure \ref{fig:Sim_plot} where all metrics are averaged over 500 data sets. We examine the mean squared error (MSE) as well as the coverage obtained from our 95\% credible intervals. To facilitate comparisons across scenarios, the MSE for each scenario is divided by the minimum MSE across all estimators for that scenario, which implies that an MSE of one is the lowest MSE possible. We consider 4 different methods: 1) a model that employs the shrinkage prior to the difference between home and neighborhood  coefficients described in Section \ref{sec: Estimation} (Shrinkage), 2) the same model without the shrinkage prior (Non-shrinkage), 3) a misspecified model that utilizes shrinkage, but also uses an incorrectly measured $\tau$ that is $0.75$ times the true mobility weights (Misspecified), and 4) a standard model that does not use mobility information at all, which is standard in air pollution studies (No mobility). We find that all models that incorporate mobility weights outperform the \textit{no mobility} approach both in terms of the MSE and coverage, which is expected given our bias results presented in Section \ref{sec:Bias}. We also see reasonably good results in terms of coverage for the models that use a misspecified $\tau$, which is expected given our bias results in Section \ref{sec:MissWeights}, but it negatively affects the MSE. Lastly, we see significant improvements in MSE by incorporating shrinkage of the two functions into the model, and the shrinkage model outperforms the model without shrinkage, even when there are moderate differences between the true home and mobility-based effects. 

\subsection{Impact of mobility misspecification in nonlinear models}\label{sec:misspecification_sim}

While Section \ref{sec:MissWeights} shows that misspecification of $\tau$ introduces only minimal bias under linear outcome models, we further conduct a simulation study here to assess the robustness of our effect estimation approach when the both mobility-related variables ($\bG$ and $\tau$) are measured with error in a more flexible nonlinear setting.

For each simulated unit, we generate bivariate exposures $\bW$ from a bivariate normal distribution with unit marginal variance and correlation 0.2. $\bG$ is generated to have correlation 0.5 with its counterpart in $\bW$. $\tau$ is drawn from a $\text{Beta}(30,10)$ distribution, and the outcome is generated as
\[
Y_i = \tau_i (W_{i1} W_{i2}) + \sin(\tau_i W_{i1})
      + (1-\tau_i) (G_{i1} G_{i2}) + 0.8 \sin\{(1-\tau_i) G_{i1}\}
      + 0.1 \tau_i (1-\tau_i) (W_{i1} G_{i1}) + \epsilon_i,
\]
where $\varepsilon_i \sim N(0,1)$. Notably, this outcome function includes nonlinear effects, interactions within each $\bW$ and $\bG$ as well as interactions between them. The true average effect was defined as the mean change in the outcome when all $\bW$ (and therefore $\bG$) were increased by 1.

\begin{table}[ht]
\centering
\caption{Estimated effects and bias under various model specifications.}
\label{tab:xgb_results}
\begin{tabular}{lcc}
\hline
\textbf{Model specification} & \textbf{Estimate} & \textbf{True value} \\
\hline
Correct $W,G,\tau$           & 1.702 & 1.720 \\
$W$ only                     & 0.925 & 1.720 \\
Noisy $\tau$ (light)         & 1.703 & 1.720 \\
Noisy $G$ (light)            & 1.704 & 1.720 \\
Noisy $\tau$ (heavy)         & 1.705 & 1.720 \\
Noisy $G$ (heavy)            & 1.630 & 1.720 \\
Noisy $G,\tau$ (light)       & 1.704 & 1.720\\
Noisy $G,\tau$ (heavy)       & 1.633 & 1.720 \\
\hline
\end{tabular}
\end{table}

We compared eight models that differed in whether $\bG$ and $\tau$ were measured accurately or with added noise. Specifically, ``light'' and ``heavy'' noise corresponded to Gaussian perturbations with standard deviations $(\sigma_G, \sigma_\tau) = (0.2, 0.05)$ and $(0.5, 0.20)$, respectively.  Each model was fit using gradient boosting (XGBoost) with squared-error loss. Table~\ref{tab:xgb_results} summarizes the results for $n = 50,000$.  The model using the correct $\bW$, $\bG$, and $\tau$ achieved nearly unbiased estimation 
of the true effect. Ignoring mobility led to substantial bias toward the null. Models with mild noise in $\bG$ and/or $\tau$ performed comparably to the correct model, suggesting that moderate measurement error in mobility has limited impact on bias. Heavy noise in $\bG$ or joint noise in $\bG$ and $\tau$ led to some degree of bias towards the null, though the magnitude of bias is much smaller than the one from the model without using mobility. Overall, these results indicate that (i) accounting for mobility is crucial for unbiased estimation, and (ii) the estimates are robust to moderate levels of mobility misspecification. Additionally, we see that the theoretical results derived in Sections \ref{sec:Bias} and \ref{sec:MissWeights} hold in a more general setting with nonlinearities and interactions.

\section{Reformulating problem into a binary treatment}
\label{app:BinaryTreatment}

Here we show that our main estimand of interest, $\omega(\boldsymbol{\Delta})$ can be estimated using inverse probability weighted estimation under an augmented data set. This provides an alternative estimation strategy that does not rely on fitting a model for the outcome given the exposures and covariates, and therefore does not rely on any distributional assumptions with respect to the outcome. Additionally, if we find similar results using this distinct modeling approach, it provides increased evidence that our main findings of the manuscript are not due to misspecification of the outcome model. 

\subsection{Methodology}

To formulate our problem as one involving inverse probability weighting and binary treatments, we extend the ideas developed in \cite{jiang2025exploring}, which focused on multiple treatment settings, to our setting that also incorporates interference. Before showing how our problem can be similarly cast as a standard binary treatment problem, we first must examine a population level version of our estimand, for which it is easier to show the connection to binary treatments. We examine a modified treatment policy similar to those explored in \cite{haneuse2013estimation} that takes the observed exposure levels and modifies them accordingly. To define our estimand, we require a function that assigns treatment (both $\boldsymbol{w}$ and $\boldsymbol{g}$ in our case) based on their observed treatment level. For simplicity throughout this section, we will refer to the home and mobility-based exposures together as $\boldsymbol{T} = [\boldsymbol{W}, \boldsymbol{G}]$. We can denote this function by
\[
q: \mathcal{T} \times \mathcal{X} \to \mathcal{T},
\]
which allows the shift in the exposures to also depend on covariates of the unit, given by $\boldsymbol{X}$. Throughout, we refer to the new value of $\boldsymbol{T}$ under the modified shift as $q(\boldsymbol{T}, \boldsymbol{X})$. With this in hand, we can define our estimand as
\begin{align*}
    \omega_p &= E(Y(q(\boldsymbol{T}, \boldsymbol{X}))) - E(Y(\boldsymbol{T})) \\
    &= E(Y(q(\boldsymbol{T}, \boldsymbol{X}))) - E(Y).
\end{align*}
This estimand quantifies the effect of shifting exposures from their natural, observed levels to those given by $q(\boldsymbol{T}, \boldsymbol{X})$. It will also be important for our subsequent discussion to show the identification formula for this potential outcome, which is given as
$$E(Y(q(\boldsymbol{T}, \boldsymbol{X}))) = \int_{(\mathcal{T} \times \mathcal{X})} m(q(\boldsymbol{t}, \boldsymbol{x}), \boldsymbol{x}) dF_{T,X}(\boldsymbol{t},\boldsymbol{x})$$
where we have that $m(\boldsymbol{t}, \boldsymbol{x})$ is equal to the conditional expectation of the outcome given exposures and covariates within our population of interest. Given the shift from the observed exposure levels, it is easy to see that this is effectively a population-level analog to our sample-level estimand given by $\omega(\boldsymbol{\Delta})$. Given this identification formula, it is clear that this estimand can be estimated by estimating the conditional expectation function $m(\boldsymbol{t}, \boldsymbol{x})$ and then integrating it over the empirical distribution of exposures and covariates found in the observed data. This is the main approach taken in the manuscript, which focuses on estimation of the conditional expectation of the outcome. However, we now show an equivalence between this estimand, and a binary treatment effect on the treated in a different population, which leads to an alternative estimation strategy based on inverse propensity score weighting. As shown in \cite{jiang2025exploring}, one can define what they refer to as an augmented population given by
$$(\boldsymbol{X}, \widetilde{\boldsymbol{T}}, Y, Z) = (\boldsymbol{X}, \boldsymbol{T}, Y, Z=0) \ \cup \ (\boldsymbol{X}, q(\boldsymbol{T}, \boldsymbol{X}), Y, Z=1),$$
where we have that $P(Z = 1) = 0.5.$  The augmented population has the same distribution of covariates and outcomes, but half the population has $Z=0$ and the same distribution of treatment as in the observed data, while half the population has $Z=1$ and the distribution of treatment given by the modified treatment levels $q(\boldsymbol{T}, \boldsymbol{X})$. Now suppose that in this augmented population, $Z$ is the treatment of interest and we therefore define potential outcomes $Y(z)$ to be the outcome we would have observed had $Z$ been set to $z$. Further, consider the set of confounders that are needed for the no unmeasured confounding assumption (when estimating the effect of $Z$ on $Y$) to hold to be $(\boldsymbol{X}, \boldsymbol{T})$. In the augmented population, the standard average treatment effect on the treated is then denoted by
$$E(Y(1) - Y(0) \mid Z = 1).$$
Crucially, it turns out that this estimand is equal to $-\omega_p$. Therefore, an alternative approach to estimating our estimand of interest, $\omega_p$, is to estimate the average treatment effect on the treated within the augmented population, which can be done using standard IPW estimation techniques that rely on estimating $P(Z=1 \mid \boldsymbol{T}, \boldsymbol{X})$. First, we show the equivalence between these two estimands:
\begin{align*}
    E(Y(1) \mid Z = 1) - E(Y(0) \mid Z = 1) &= E(Y \mid Z = 1) - E(Y(0) \mid Z = 1) \\
    &= E(Y) - E(Y(0) \mid Z = 1) \\
    &= E(Y) - E[E(Y(0) \mid \boldsymbol{T}, \boldsymbol{X}, Z = 1) \mid Z=1] \\ &= E(Y) - E[E(Y(0) \mid \boldsymbol{T}, \boldsymbol{X}, Z = 0) \mid Z=1] \\ &= E(Y) - E[E(Y \mid \boldsymbol{T}, \boldsymbol{X}, Z = 0) \mid Z=1] \\
    &= E(Y) - \int_{(\mathcal{T} \times \mathcal{X})} m(\boldsymbol{t}, \boldsymbol{x}) dF_{T,X \mid Z=1}(\boldsymbol{t},\boldsymbol{x}) \\
    &= E(Y) - \int_{(\mathcal{T} \times \mathcal{X})} m(q(\boldsymbol{t},\boldsymbol{x}), \boldsymbol{x}) dF_{T,X}(\boldsymbol{t},\boldsymbol{x}) \\
    &= E(Y) - E(Y(q(\boldsymbol{T}, \boldsymbol{X}))) \\
    &= -\omega_p.
\end{align*}
Here we used that $m(\boldsymbol{T}, \boldsymbol{X})$ is the conditional expectation of the outcome in the original population, which corresponds to the $Z=0$ population within the augmented population. Now that we have shown this equivalence, we can state explicitly how to estimate $\omega_p$ using IPW weighting within the augmented population. Specifically, we create a new data set that has a sample size of $2n$, where each data point has exactly two copies: one with the original exposure and one with the modified exposure. For simplicity we can assume that the first $n$ observations are the original observations, which all have $Z_i = 0$ for $i=1, \dots, n$. The remaining observations have $Z_i = 1$ for $i=n+1, \dots, 2n$. Note also that the exposure values for these observations are the modified values in the sense that $\boldsymbol{T}_i = q(\boldsymbol{T}_{i-n}, \boldsymbol{X}_{i})$ for $i=n+1, \dots, 2n$. Given this modified data set, we then simply use the IPW estimator for the average treatment effect on the treated, which in our setting is given by:
$$\frac{1}{n} \sum_{i=1}^{2n} \left\{ Z_i Y_i - \frac{(1 - Z_i) e(\boldsymbol{T}_i,\boldsymbol{X}_i) Y_i}{1 - e(\boldsymbol{T}_i,\boldsymbol{X}_i)} \right\},$$
where we let $e(\boldsymbol{T}_i,\boldsymbol{X}_i)$ be equal to the propensity score, $P(Z=1 \mid \boldsymbol{T}, \boldsymbol{X})$, which can be estimated using any approach for binary outcome regression applied to the augmented data set. Note that this relies only on estimation of $P(Z=1 \mid \boldsymbol{T}, \boldsymbol{X})$ and does not require modeling of the conditional expectation of the outcome, and therefore provides a distinct estimation strategy for our estimand of interest than the one used in the manuscript.

\subsection{Results for the effect of air pollution mixture}

\begin{figure}
    \centering
    \includegraphics[width=\linewidth]{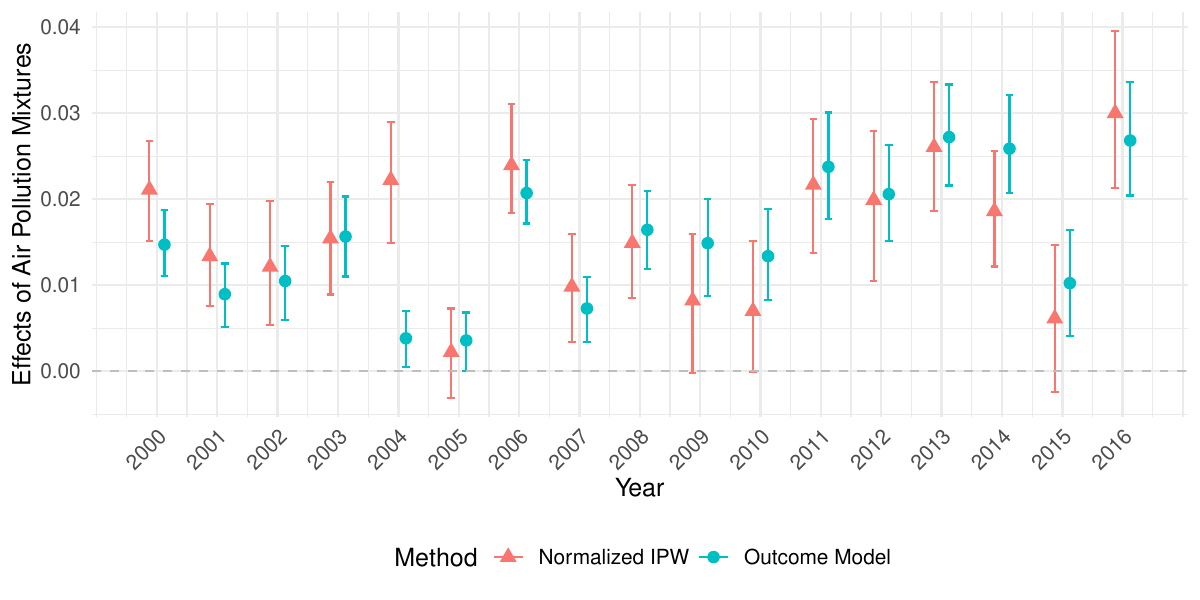}
    \caption{Effects of the air pollution mixture comparing the approach based on outcome modeling and the one based on IPW estimation. Note that both approaches incorporate mobility information. }
    \label{fig:ResultsIPWmixture}
\end{figure}

Here we compare the results using the approach described in the manuscript with those described above using inverse probability weighted estimation. The results for estimating a shift in the entire exposure mixture can be found in Figure \ref{fig:ResultsIPWmixture}. The main takeaway from the results is that we find consistent results between the IPW based approach and the outcome modeling approach. While the numeric results are not identical between these two distinct approaches in any particular year, they tend to be quite close to each other, and present a very similar story across all years in terms of the magnitude and significance of the estimated effects. 

\section{Sensitivity to unmeasured confounding}\label{app:unmeasuredconfounding}

Throughout our analyses, we assumed that $\boldsymbol{X}$ contained all of the relevant confounders of the exposure-outcome associations of interest. If there are unmeasured confounders that affect both the exposures and outcome, this could lead to biased results of the causal effects of interest. Therefore, in this section, we develop a sensitivity analysis framework to assess the robustness of our findings to unmeasured confounders. There are a number of different approaches in causal inference to performing sensitivity analysis, though their overarching goal is to bound the bias in terms of interpretable sensitivity parameters that can be reasoned about, and then assess how large these parameters must become to attenuate any significant causal effects towards zero. These sensitivity parameters are commonly framed in terms of partial $R^2$ values \citep{imbens2003sensitivity, chernozhukov2021long}, relative risks \citep{ding2016sensitivity}, or parameters of regression models \citep{rosenbaum1983assessing, mccandless2007bayesian}, each of which control the strength of association between the unmeasured variable and the exposures or outcome. Most of these approaches, however, are focused on settings with a single exposure and a single outcome, whereas we have a vector of exposures, along with their neighboring exposure counterparts. The main complication that arises due to the multivariate nature of our exposure vector is that the bias of causal effects of interest is driven by a large number of sensitivity parameters, which makes them less interpretable and harder to reason about. For this reason, we use the observed covariates to guide our sensitivity analysis and reduce the number of sensitivity parameters significantly. 

To formalize our discussion of sensitivity analysis, we assume the following data generating models for the outcomes, exposures, and unmeasured variable.
\begin{align*}
    Y &= f(\boldsymbol{W}, \boldsymbol{G}, \boldsymbol{X}) + \gamma U + \epsilon_y \\
    \boldsymbol{T} &= h(\boldsymbol{X}) + \boldsymbol{\delta}U + \epsilon_t \\
    U &\sim \mathcal{N}(0, 1),
\end{align*}
where $\epsilon_y$ and $\epsilon_t$ are i.i.d residuals with mean zero and variances given by $\sigma^2_{y|t,x,u}$ and $\boldsymbol{\Sigma}_{t|x,u}$. Here we adopt the notation that $\boldsymbol{T} = [\boldsymbol{W}, \boldsymbol{G}]$ contains our full set of exposures. This makes no assumptions about how the observed variables are related to each other, but assumes that the unmeasured variable affects the exposures and outcomes in an additive way that does not depend on the values of the observed covariates, or the values of the exposures in the case of the outcome model. Further suppose that the causal effect we are trying to estimate corresponds to shifting from the observed exposure levels $\boldsymbol{T}_i$ to the shifted levels given by $\boldsymbol{T}_i + \boldsymbol{\Delta}_i$. Lastly, let the average exposure shift be denoted by $\overline{\boldsymbol{\Delta}} = (1/n) \sum_{i=1}^n \boldsymbol{\Delta}_i$. The bias one obtains when estimating the causal effect by ignoring the presence of $U$ is given by
$$\text{bias} = \gamma (\boldsymbol{\delta}^T \boldsymbol{\Sigma}_{t|x}^{-1} \overline{\boldsymbol{\Delta}}),$$
where $\boldsymbol{\Sigma}_{t|x} = \text{Var}(\boldsymbol{T} \mid \boldsymbol{X})$ is a fully identifiable quantity. This shows that the bias depends on two sensitivity parameters $\gamma$ and $\boldsymbol{\delta}$, and one can see from this expression that if either $\gamma = 0$ or $\boldsymbol{\delta} = \boldsymbol{0}$ then the unmeasured variable is not a confounder and there is no confounding bias. Given this expression for bias, we can take a Bayesian approach to sensitivity analysis \citep{mccandless2007bayesian, zheng2021bayesian} where we place prior distributions over the important parameters $\gamma$ and $\boldsymbol{\delta}$, which in turn provides a posterior distribution for the bias of the causal effect of interest. Coupled with our posterior distribution of our causal estimand, this provides a posterior distribution of the causal effect that allows for the potential violations of the no unmeasured confounding assumption. The degree of these violations is given by the prior distributions for $\gamma$ and $\boldsymbol{\delta}$, which we detail now.

We take a similar approach as that used in \cite{mccandless2007bayesian} to specifying prior distributions for these unidentifiable parameters, however, we utilize the observed covariates to find reasonable prior specifications, which is common in the causal inference literature on sensitivity analysis \citep{cinelli2020making}. To begin, we start with $\gamma$, and we utilize a result relating this regression parameter with $R^2_{U \sim Y \mid T,X}$ and  $R^2_{U \sim T \mid X}$. These correspond to the partial $R^2$ between $U$ and $Y$ given $\boldsymbol{T}$ and $\boldsymbol{X}$ and the partial $R^2$ between $U$ and $\boldsymbol{T}$ given $\boldsymbol{X}$. Specifically, we have that
$$|\gamma| = \sqrt{\frac{\sigma^2_{y|t,x} R^2_{U \sim Y \mid T,X}}{\text{Var}(U \mid \boldsymbol{T}, \boldsymbol{X})}} = \sqrt{\frac{\sigma^2_{y|t,x} R^2_{U \sim Y \mid T,X}}{1 - R^2_{U \sim T | X}}}$$
where $\sigma^2_{y|t,x} = \text{Var}(Y \mid \boldsymbol{T}, \boldsymbol{X})$. Note that the second equality holds because $\text{Var}(U) = 1$ and $U$ is independent of $\boldsymbol{X}$. This result is useful because partial R-squared values are generally easier to reason about, and because there is a rich literature about how to use the observed covariates to reason about them. We adopt the formal benchmarking framework of \cite{cinelli2020making} where we assume that the association between the unmeasured variable and the outcome (given exposures and the remaining covariates) is as strong as the association between the strongest observed covariate and the outcome (given exposures and the remaining covariates). Specifically, we set 
$$k_y = \frac{R^2_{U \sim Y \mid T,X_{-j}}}{R^2_{X_j \sim Y \mid T,X_{-j}}},$$
which in turn provides a bound for $R^2_{U \sim Y \mid T,X}$ that can be found in \cite{cinelli2020making}. 
Throughout, we set $k_y = 1$ and therefore assume that the unmeasured variable is as strongly associated with the outcome as the strongest observed covariate. A similar process occurs for benchmarking $R^2_{U \sim T | X}$ using observed covariates, and we again assume that the association between the unobserved variable and exposures is equally as strong as the association between the strongest observed covariate and the exposures. This quantity is slightly more complicated to benchmark since $\boldsymbol{T}$ is a vector of exposures and therefore the results from \cite{cinelli2020making} do not immediately apply. To address this, one can use the result stating that
$$R^2_{U \sim T | X} = \boldsymbol{r}^T \boldsymbol{S} \boldsymbol{R}^{-1} \boldsymbol{S} \boldsymbol{r},$$
where $\boldsymbol{r}$ is the vector of partial R-squared values of the form $R^2_{U \sim T_j | X}$, $\boldsymbol{S}$ is a diagonal matrix of the signs of the corresponding partial correlations, and $\boldsymbol{R}$ is the conditional correlation of $\boldsymbol{T}$ given $\boldsymbol{X}$. One can use the formal benchmarking techniques from \cite{cinelli2020making} for each element $R^2_{U \sim T_j | X}$, which provides a benchmark value for the quantity of interest, $R^2_{U \sim T | X}$. Due to the formula above that relates $\gamma$ with these partial R-squared values, this provides a value for $\gamma$ that is benchmarked from the observed covariates, which we refer to as $\gamma_{max}$. For our sensitivity analysis, we specify a uniform prior on the interval $[-\gamma_{max}, \gamma_{max}]$ to allow for violations of unmeasured confounding as strong as those provided by this benchmark value, and in either direction. 

Adopting the same strategy does not work for the exposure parameters because the bias can not be expressed directly in terms of partial $R^2$ values, and the bias further depends on the interplay between $\boldsymbol{\delta}$ and the covariance matrix of the exposures given the observed covariates. For these reasons, we directly specify a prior on $\boldsymbol{\delta}$, however, this is a difficult task as well given the multivariate nature of our exposure vector. We assume a multivariate normal distribution for $\boldsymbol{\delta}$ and use the observed covariates to provide reasonable values for the mean and covariance terms. Specifically, we fit a multivariate regression model with $\boldsymbol{T}$ as the outcome, and $\boldsymbol{X}$ as the predictors:
$$E(\boldsymbol{T} \mid \boldsymbol{X}) = \boldsymbol{\widetilde{\delta}}_0 + \sum_{j=1}^p X_j \boldsymbol{\widetilde{\delta}}_j.$$
We then identify the covariate $j^*$ that changes the estimate of the causal effect the most when it is removed from estimation. For this covariate, we find the point estimate and covariance matrix of $\boldsymbol{\widetilde{\delta}}_{j^*}$ from the linear model fit, and we use these as the mean and covariance matrix of the prior distribution for $\boldsymbol{\delta}$. Similar to the outcome parameter $\gamma$, this process utilizes the observed covariates to guide the sensitivity analysis and assumes that the unmeasured variable is as strongly associated with the exposures as the strongest observed covariate. 

\subsection{Sensitivity of the estimated health effects of air pollution}

\begin{figure}
\centering
    \includegraphics[width=\linewidth]{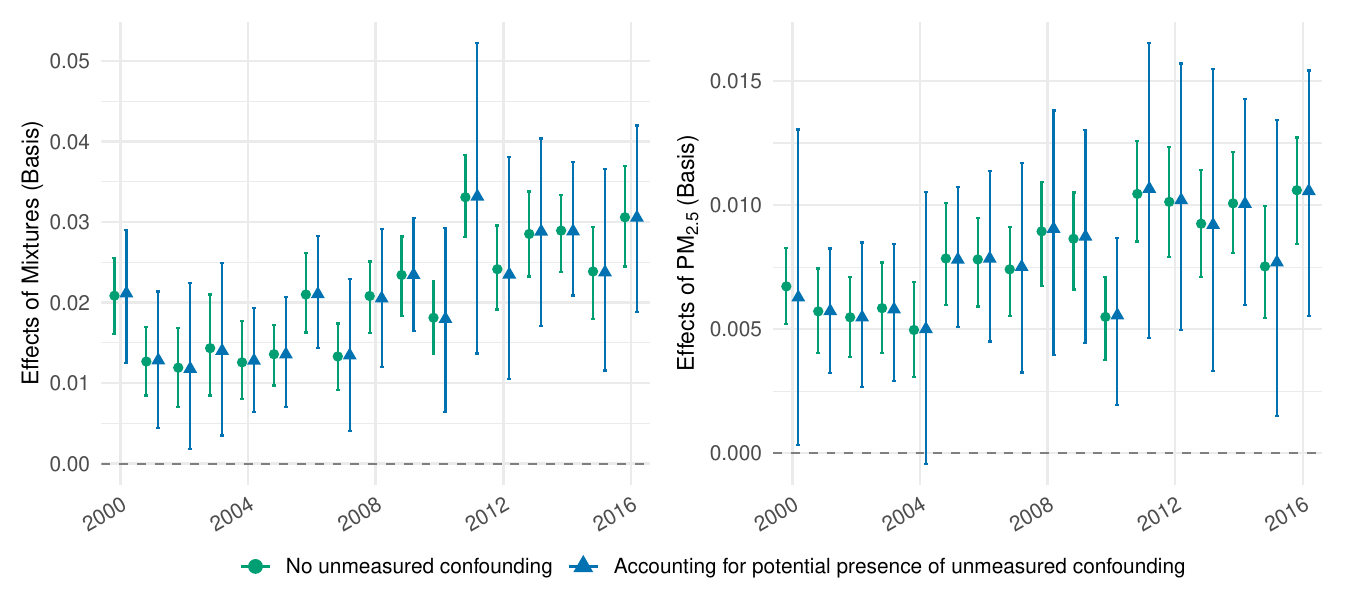}
    \caption{Sensitivity analysis for unmeasured confounding using the basis expansion estimator. The left panel corresponds to the effect of the air pollution mixture, while the right panel corresponds to the effect of \pmtpf.}
        \label{fig:SensitivityUnmeasured}
\end{figure}

\begin{figure}
\centering
    \includegraphics[width=\linewidth]{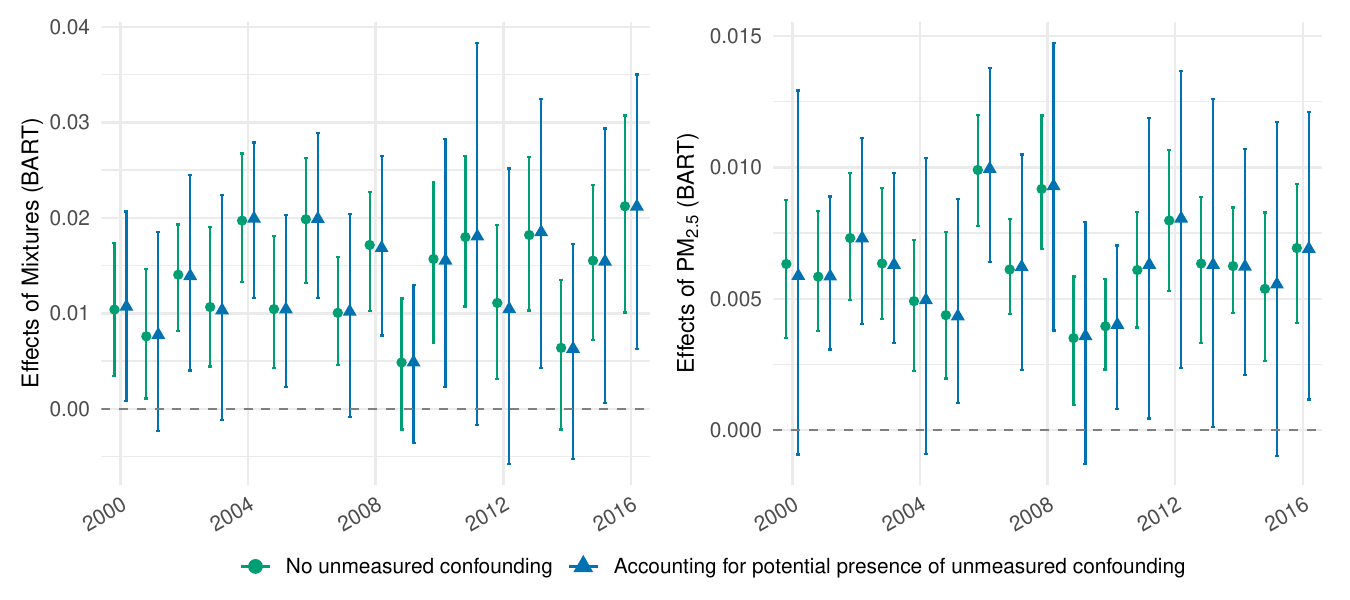}
    \caption{Sensitivity analysis for unmeasured confounding using the BART estimator. The left panel corresponds to the effect of the air pollution mixture, while the right panel corresponds to the effect of \pmtpf.}
        \label{fig:SensitivityUnmeasured_BART}
\end{figure}

We now apply this process to the main estimand discussed in our manuscript, which involves shifting all components of the air pollution mixture. As a complementary analysis, we also do sensitivity analysis for analyzing the univariate \pmtpfs effect as well. These results can be found in Figures \ref{fig:SensitivityUnmeasured} and \ref{fig:SensitivityUnmeasured_BART}, which show the sensitivity analysis results for the basis expansion and BART estimators, respectively. For each estimator, these figures show the original 95\% credible intervals that assume no unmeasured confounding, along with the 95\% credible intervals obtained while accounting for the potential presence of unmeasured confounding that is described above.  As expected, the intervals when accounting for unmeasured confounding are substantially wider than those that assume no unmeasured confounding, as they allow for the possibility of confounding bias in our causal estimates. Importantly, however, for the basis expansion estimator nearly all of the 95\% credible intervals remain entirely above zero providing robust evidence in support of the presence of a causal effect of the air pollution mixture on mortality. The point estimates from the BART estimator are generally closer to zero, and therefore the intervals allowing for unmeasured confounding are more likely to contain zero. Despite this, the majority of the intervals from the BART estimator are still exclusively above zero even after allowing for the presence of unmeasured confounding, which shows a moderate degree of robustness to unmeasured confounding. Note that this analysis does not rule out entirely the possibility that unmeasured confounding could be driving the causal effects seen in our analysis. Rather, it shows that an unmeasured confounder would have to be substantially stronger than the strongest of the observed confounders in order to fully explain away the significant causal effects seen. 


\section{Accounting for spatial correlation}
\label{app:SpatialICAR}

Throughout our main analyses, we assumed that the outcomes for each unit were conditionally (on exposures and covariates) independent of each other. It is possible, however, that there exists residual spatial dependence, particularly among nearby zip codes. If such spatial correlation in the outcomes is present and our models ignore it, our inferences may be invalid, and of particular concern is that we may underestimate the degree of uncertainty in our estimates. To address this issue, we extend our models here to incorporate spatial random effects that have an intrinsic conditional autoregressive (ICAR) prior distribution that encourages similarity across neighboring zip codes with respect to the random effects, thereby inducing spatial correlation among neighboring units \citep{besag1974spatial, hodges2003precision, tessema2023systematic}. We briefly review the ICAR prior distribution within the context of our model specification, and then explore the extent to which accounting for spatial correlation changes results in our application of the health effects of air pollution mixtures. 

\begin{figure}
    \centering
    \includegraphics[width=\linewidth]{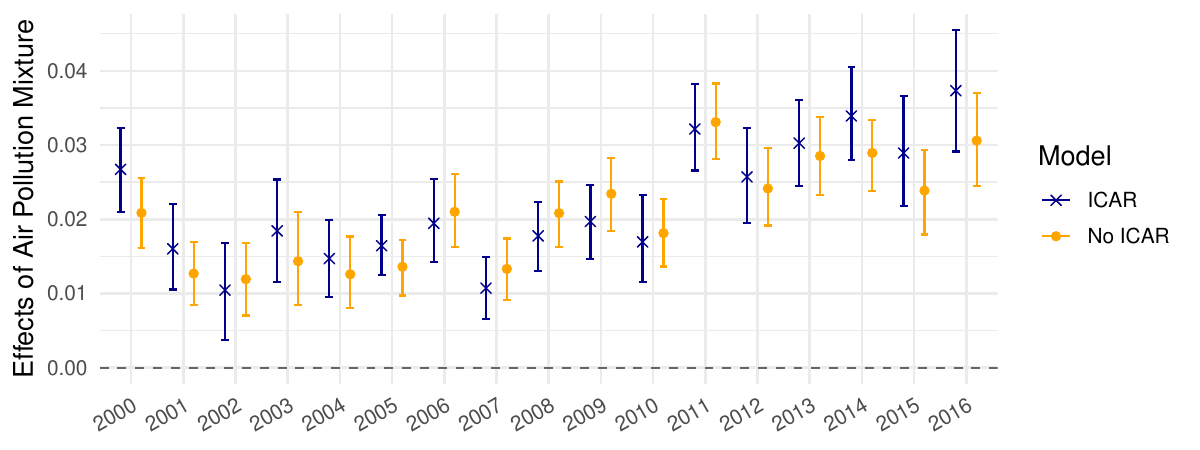}
    \caption{Results from the main analysis looking at the effect of the air pollution mixture on mortality when an ICAR prior is placed on random effects to account for spatial dependence. }
    \label{fig:ICARresults}
\end{figure}

Throughout, we will now be assuming the following model
$$Y_i \mid \boldsymbol{W} = \boldsymbol{w}_i, \boldsymbol{G} = \boldsymbol{g}_i, \boldsymbol{X} = \boldsymbol{x}_i \sim \mathcal{N}(u_i + \mu(\boldsymbol{w}_i, \boldsymbol{g}_i, \boldsymbol{x}_i), \sigma^2),$$
where $\mu(\boldsymbol{w}_i, \boldsymbol{g}_i, \boldsymbol{x}_i)$ can be any functional form explored in the manuscript such as the additive expansion that encourages similarity between the effects of $\boldsymbol{W}$ and $\boldsymbol{G}$, or any of the BART based models explored. We focus here on the prior distribution for $u_i$, which will encourage similarity between nearby zip codes. If we let $\mathcal{N}_i$ be the indices corresponding to neighbors of zip code $i$ and $n_i$ be the number of neighbors, then the prior distribution is commonly written under it's conditional specification given by
$$u_i \mid u_{-j} \sim \mathcal{N} \bigg(\frac{1}{n_i} \sum_{j \in \mathcal{N}_i} u_j, \frac{\sigma^2_u}{n_i} \bigg).$$
Clearly, this encourages the random effects of neighboring zip codes to be similar as the random effects are centered at the mean of the random effects for the neighboring areas. This leads to a joint distribution for the random effects that is given by
$$P(\boldsymbol{u}) \propto \text{exp} \bigg(-\frac{1}{2 \sigma^2_u} \sum_{i \sim j} (u_i - u_j)^2 \bigg),$$
where the notation $i \sim j$ denotes a summation over all neighboring pairs of zip codes. To make this joint distribution proper, we enforce a standard sum to zero constraint that $\sum_i u_i = 0$. We utilize this random effect structure within the same model from which we obtain our main results in the manuscript, and the results can be found in Figure \ref{fig:ICARresults}. We see that the results are nearly indistinguishable from those seen in the manuscript, which points to the fact that incorporating spatial correlation does not alter the point estimates or widths of the credible intervals in a material way. 

\section{Additional Medicare analysis results}\label{app:pm_analysis}

\subsection{Exploring the impact of mobility}
\label{app:ssec:mobility}

\begin{figure}[t]
    \centering
    \includegraphics[width=0.75\linewidth]{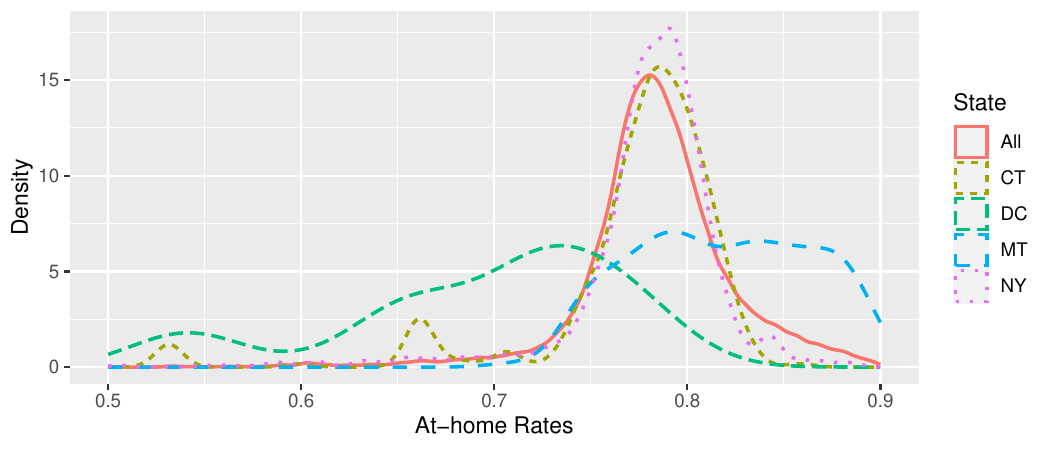}
    \caption{Density plots of the proportion of time people stay in their residential zip code $(\tau_i)$ in 2019 across geographic regions}
    \label{fig:tau_state}
\end{figure}

\begin{figure}[t]
    \centering
    \includegraphics[width=0.8\linewidth]{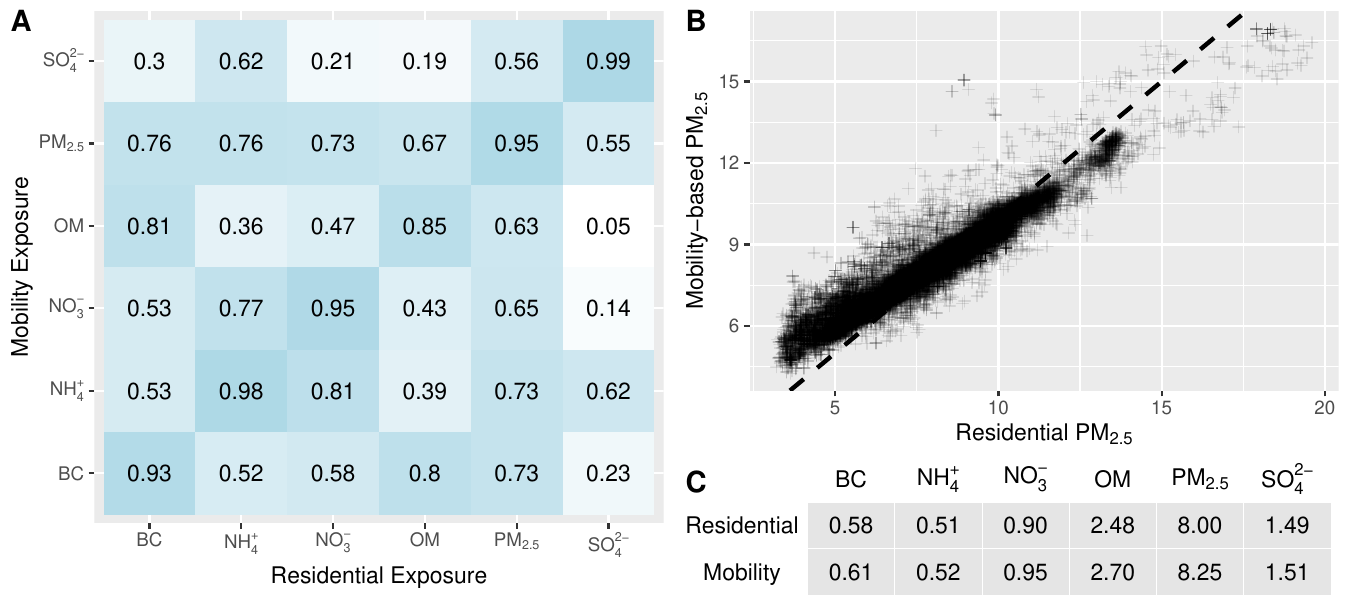}
    \caption{Summary of residential and mobility exposures in 2013. (A) Correlation between residential and mobility exposures; (B) Scatter plot of residential and mobility-based \pmtpf; (C) Average exposure levels by pollutant.}
    \label{fig:home_mob}
\end{figure}

We can first look at exposure levels that incorporate mobility to assess the potential impact of mobility on subsequent causal effect estimates. Figure \ref{fig:tau_state} shows the proportion of time individuals stayed in their residential zip code in 2019 across the entire United States and for a few selected states. On average, individuals spend approximately 78\% of their time in their residential zip code, though there is some variation across states. Of equal importance, is the exposure levels at the locations that people are traveling to during daily mobility. Figure \ref{fig:home_mob} shows a number of metrics that compare residential and mobility-based exposure to air pollution. We see in Panel (A) that there is a very high correlation between residential and mobility-based exposures for all pollutants considered. Panel (B) shows this relationship for \pmtpfs specifically, where we see that individuals in low \pmtpfs zip codes tend to travel to areas with higher pollution levels, and those in high pollution areas tend to travel to areas with lower \pmtpf. Similar trends were seen for all pollutants considered in all years. Panel (C) shows that mobility-based exposure to air pollution is higher for all pollutants than their corresponding residential exposure level. This could be due to individuals living in suburban areas with lower pollution levels commuting to urban centers with higher pollution levels, where workplaces and commercial activities are concentrated.

Overall, the mobility data shows that people do travel outside of their residential zip code to areas with different levels of pollution, and that typically this is to higher levels of pollution than what they are exposed to in their residential area. The correlation between the residential and mobility-based exposures is quite high, which suggests that accounting for mobility won't drastically impact causal effect estimates of the impact of air pollution on mortality. 

\subsection{Positivity assessment}
\label{app:ssec:positivity}

Here we empirically assess the positivity assumption in our Medicare application. There is not a well-established method for determining whether positivity holds in the context of multivariate, continuous treatments. In our setting, we elect to first estimate the generalized propensity score \citep{imai2004causal}, which in our case is the conditional density of $(\bW, \bG)$ given $\bX$, denoted by $f_{\bW,\bG|\bX}(\cdot)$. We assume that this density follows a normal distribution with a mean that is a cubic polynomial function of the covariates. After estimating this density, for every observation we calculate two distinct values from this estimated density:
\begin{enumerate}
    \item $f_{1i} = \widehat{f}_{\bW,\bG|\bX}(\bW_i,\bG_i \vert \bX_i)$
    \item $f_{2i} = \widehat{f}_{\bW,\bG|\bX}((\boldsymbol{W}_i + \boldsymbol{\Delta}_{wi}, \boldsymbol{G}_i+ \boldsymbol{\Delta}_{gi}) | \bX = \bX_i)$
\end{enumerate}
We know that there is positive support at $(\bW_i, \bG_i, \bX_i)$ due to the existence of observation $i$, but it is not guaranteed that there is positive support at $(\boldsymbol{W}_i + \boldsymbol{\Delta}_{wi}, \boldsymbol{G}_i+ \boldsymbol{\Delta}_{gi}, \bX_i)$. To investigate the positivity violation we can look at the ratio of these two quantities, given by $r_i = f_{2i} / f_{1i}$. If the positivity assumption holds, we would expect this ratio to be close to 1 on average, whereas if positivity does not hold in our data, we would expect to see ratios that are generally smaller than 1. 

For our multivariate analysis of the air pollution mixture, we find that the mean of $r_i$ across the sample is exactly 1, and the 0.05 and 0.95 quantiles of the distribution of $r_i$ are 0.65 and 1.45. Overall, this suggests that positivity violations are not a big concern in our analysis, which is to be expected given the choice of estimand that shifts from the observed exposure level by only a small amount. Note that while we performed this exercise for the entire air pollution mixture where positivity is of a bigger concern, we also performed this exercise where our only exposure is PM$_{2.5}$. We again find that the mean of $r_i$ is equal to 1 and the 0.05 and 0.95 quantiles are 0.83 and 1.18, respectively. Again, this suggests that positivity holds in our sample at least approximately. 

\subsection{Effects of total \texorpdfstring{PM$_{2.5}$}{PM2.5}}

\begin{figure}
    \centering
    \includegraphics[width=\linewidth]{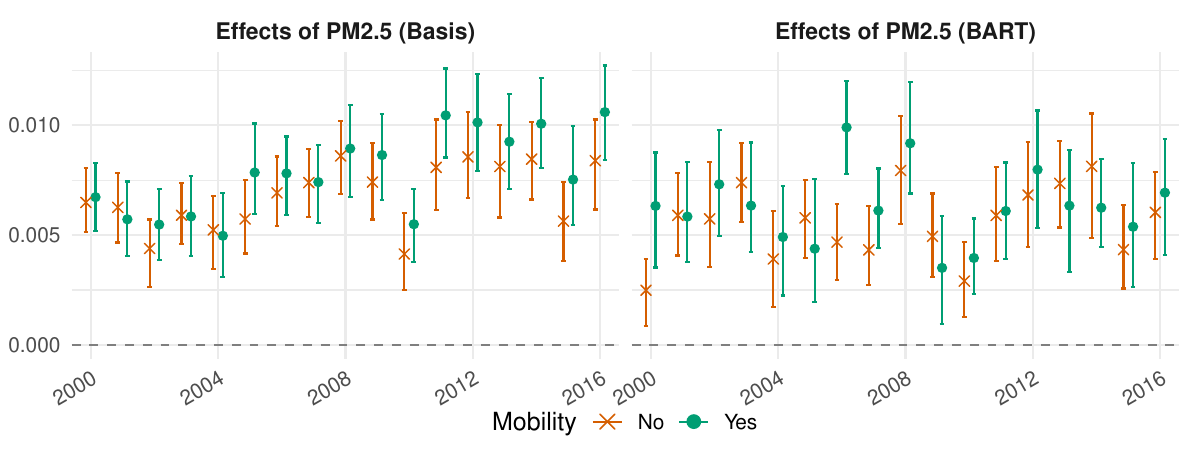}
    \caption{Estimated effects of \pmtpf.}
    \label{fig:pm_main}
\end{figure}

Here we present the results for the single exposure analyses with total \pmtpf. Recall that our goal is to estimate the effect of shifting \pmtpfs by 0.05 standard deviations across all units. Figure \ref{fig:pm_main} shows the estimated causal effect of \pmtpfs on mortality using the basis expansion and BART models for each year separately. Consistent with the multivariate exposure analysis, we observe statistically significant adverse effects of increased \pmtpfs for both the basis expansion and BART models, with the model accounting for mobility yielding slightly larger effects.

\subsection{Sensitivity to mobility misspecification}
\label{app:ssec:MisspecifyMobility}
We conducted a sensitivity analysis to evaluate how uncertainty in the mobility weights ($\tau$) and mobility-based exposures ($\bG$) could affect the estimated treatment effects presented in Section \ref{sec: data analysis}.

For each unit, we generate noisy versions of $\tau_i$ and $\bG_i$ from Beta distributions that preserve their means while inflating their uncertainty. Specifically, for each original $\tau_i$, we defined $$\widetilde\tau_i\sim\text{Beta}(\alpha_i,\beta_i),\qquad \alpha_i=\tau_i\phi, ~\beta_i=(1-\tau_i)\phi$$
where $\phi$ controls how tightly the distribution is concentrated around the mean $\tau_i$. Smaller values of $\phi$ produce strong noise. We parametrize $\phi= 200 - 180L$ for $L\in(0,1]$ so that $L$ directly represents the noise level. When $L=0$, we set $\widetilde\tau_i=\tau_i$, representing a perfectly measured $\tau_i$. The variance of $\widetilde\tau_i$ increases with $L$, and when $L=1$ and $\tau_i=0.5$, the standard deviation of $\widetilde\tau_i$ is approximately 0.1, and roughly 95\% of the draws lie between 0.3 to 0.7, representing substantial measurement error. We apply  the same Beta resampling strategy to generate noisy mobility-based exposures $\widetilde\bG_i$. We first rescale each component of $\bG_i$ to the unit interval by setting $G^\dag _{ij} = (G_{ij} - \min(\bG_{.j}))/(\max(\bG_{.j})-\min(\bG_{.j}))$ and then we draw
$$\widetilde G^\dag_{ij}\sim\text{Beta}(\alpha_i,\beta_i),\qquad \alpha_i=G^\dag _{ij}\phi, ~\beta_i=(1-G^\dag _{ij})\phi.$$
Finally, $\widetilde G^\dag_{ij}$ are mapped back to the original scale of each exposure component.

\begin{figure}
    \centering
    \includegraphics[width=0.8\linewidth]{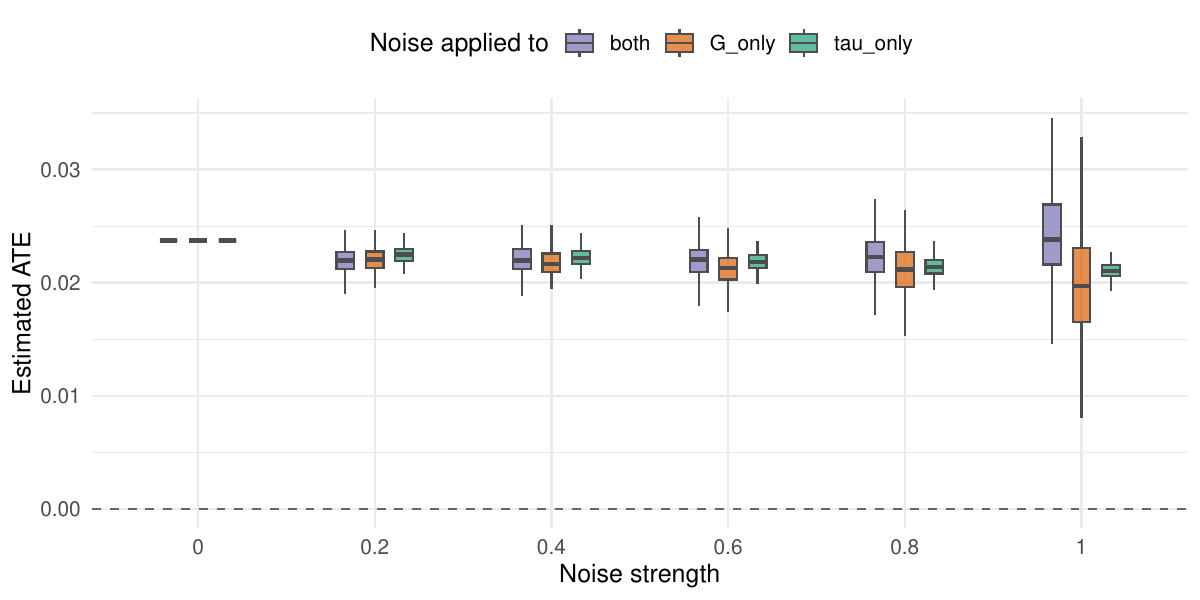}
    \caption{Estimated effects of air pollution mixtures with misspecified mobility.}
    \label{fig:sensitivity2}
\end{figure}
We considered three misspecified models in which noise was added to either $\bG$, $\tau$, or both, with varying noise levels. Figure \ref{fig:sensitivity2} presents box plots of the estimated sample average treatment effect based on 200 replicates for each noise level. When either $\bG$ or $\tau$ is misspecified, the estimated effects are biased toward the null, and this bias increases as the noise level grows. Overall though, the magnitude of the bias remains very small relative to the estimated values and shows that the effects are robust to such noise. When noise is added to both mobility-related variables, the bias is still quite small, though is not necessarily towards the null of no causal effects. Overall, the consistency of the estimates across different mobility misspecification scenarios supports the robustness of our findings on the adverse effects of air pollution mixtures.

\end{document}